\documentclass[screen, acmlarge, manuscript]{acmart}
\setcitestyle{numbers}

\usepackage{enumerate}
\usepackage{caption}
\usepackage{subcaption}
\usepackage{graphicx}
\AtBeginDocument{%
  \providecommand\BibTeX{{%
    \normalfont B\kern-0.5em{\scshape i\kern-0.25em b}\kern-0.8em\TeX}}}

\setcopyright{none}
\renewcommand\footnotetextcopyrightpermission[1]{}
\usepackage{xspace}
\usepackage{xcolor}
\usepackage{multirow}
\usepackage{soul} 
\usepackage{enumitem}
\usepackage{booktabs}
\usepackage{rotating}

\newcommand{\pro}{{\small \sf PRO}\xspace}
\newcommand{\con}{{\small \sf CON}\xspace}
\newcommand{\touche}{Touch{\'e}\xspace}

\title[Towards Detecting and Mitigating Cognitive Bias in Spoken Conversational Search]{Towards Detecting and Mitigating Cognitive Bias in Spoken Conversational Search}

\author{Kaixin Ji}
\authornote{Equal contributions.}
\orcid{0000-0002-4679-4526}
\affiliation{
\institution{RMIT University}
  \city{Melbourne}
  \country{Australia}
}
\email{kaixin.ji@student.rmit.edu.au}

\author{Sachin Pathiyan Cherumanal}
\authornotemark[1]
\orcid{0000-0001-9982-3944}
\affiliation{
\institution{RMIT University}
  \city{Melbourne}
  \country{Australia}
}
\email{sachin.pathiyan.cherumanal@student.rmit.edu.au}

\author{Johanne R.~Trippas} 
\orcid{0000-0002-7801-0239}
\affiliation{
\institution{RMIT University}
  \city{Melbourne}
  \country{Australia}
}
\email{j.trippas@rmit.edu.au}

\author{Danula Hettiachchi}
\orcid{0000-0003-3875-5727}
\affiliation{
\institution{RMIT University}
  \city{Melbourne}
  \country{Australia}
}
\email{danula.hettiachchi@rmit.edu.au}

\author{Flora D. Salim}
\orcid{0000-0002-1237-1664}
\affiliation{%
  \institution{The University of New South Wales} 
  \city{Sydney}
  \country{Australia}
}
\email{flora.salim@unsw.edu.au}

\author{Falk Scholer}
\orcid{0000-0001-9094-0810}
\affiliation{
\institution{RMIT University}
  \city{Melbourne}
  \country{Australia}
}
\email{falk.scholer@rmit.edu.au}

\author{Damiano Spina} 
\orcid{0000-0001-9913-433X}
\affiliation{
\institution{RMIT University}
  \city{Melbourne}
  \country{Australia}
}
\email{damiano.spina@rmit.edu.au}

\renewcommand{\shortauthors}{Ji~et~al.}

\begin{document}

\begin{abstract}
Instruments such as eye-tracking devices have contributed to understanding how users interact with screen-based search engines. However, user-system interactions in audio-only channels -- as is the case for Spoken Conversational Search (SCS) -- are harder to characterize, given the lack of instruments to effectively and precisely capture interactions.
Furthermore, in this era of information overload, cognitive bias can significantly impact how we seek and consume information -- especially in the context of controversial topics or multiple viewpoints.
This paper draws upon insights from multiple disciplines (including information seeking, psychology, cognitive science, and wearable sensors) to provoke novel conversations in the community.  
To this end, we discuss future opportunities and propose a framework including multimodal instruments and methods for experimental designs and settings. We demonstrate preliminary results as an example. We also outline the challenges and offer suggestions for adopting this multimodal approach, including ethical considerations, to assist future researchers and practitioners in exploring cognitive biases in SCS.







\end{abstract}

\ccsdesc[500]{Human-centered computing~Empirical studies in ubiquitous and mobile computing}
\ccsdesc[500]{Information systems~Users and interactive retrieval}

\keywords{Cognitive Bias, Spoken Conversational Search, Information Seeking, Physiological Signals, Wearable Sensors, Experimental Design}

\maketitle

\section{Introduction}
\label{sec:intro}

The integration of generative AI into search engines, such as ChatGPT and Bing Chat, is a testament to the shift of traditional information search from ``ten blue links'' to Conversational Search, which follows a question-answering paradigm.
Although such generative AI technologies are currently primarily text-based, with the growing use of intelligent assistants (e.g., Google Assistant, Amazon Alexa, Apple Siri) and the pursuit of natural human-computer interaction, fully voice-based information interaction is just around the corner.\footnote{\url{https://openai.com/blog/chatgpt-can-now-see-hear-and-speak} [Accessed: 9 Feb 2024]}
Such Spoken Conversational Search (SCS) can address complex information needs for a large population with limited access, including the visually impaired, low-literacy communities, and also the sighted users in scenarios that preclude reading, such as while driving, cooking, or exercising. 
Yet, delivering user-friendly and valuable responses from such systems via a voice channel is not straightforward.

Search systems have emerged as a primary source of information. Highlighted by recent research on information bias~\cite{biasreview2021}, it is crucial for these systems to not only provide relevant information but also present a broad spectrum of perspectives, curating and offering content that is accurate, relevant, and reflective of different viewpoints. This curated approach may aid in promoting a more informed and balanced understanding among users, and mitigating potential ``echo chambers''. Besides, researchers have raised concerns about biases in personalized informatics \cite{Yfantidou2023bias}, or Conversational Search \cite{sharma2024echo}. In addition, search engines play a crucial role beyond convenience: given the human brain's limited capacity to process and analyze information, together with cognitive biases that influence how we perceive and interpret data, they act as mediators of knowledge, tasked with ethically curating content to help users overcome these limitations and biases.
Throughout this paper, we adopt the following definition of cognitive biases and keep it in mind during our discussions.

\begin{quote}
{``Cognitive biases are systematic errors in judgment and naturally occurring tendencies that skew information processes, due to limitations in cognitive, motivational, or environmental factors, which lead to sub-optimal or fundamentally wrong outcomes''~\cite {wilke2012cognitive}.} 
\end{quote}

In today's digital age, users are commonly bombarded with vast amounts of information, which may lead to stress, confusion, and reduced productivity \cite{yunkaporta2023right}.\footnote{This is termed as ``Information overload'', a situation occurs when an abundance of relevant information is available that becomes a hindrance rather than a help~\cite{yunkaporta2023right}.} 
Effectively navigating through the information requires strategies such as filtering, prioritizing, and organizing, where cognitive biases provide shortcuts. But they are also easily ``weaponized''~\cite{alaofi2022query} to manipulate public opinion~\cite{spina2023human,yunkaporta2023right}.
For example, politicians strategically publicize specific news stories while suppressing others, leveraging the \textit{Availability Heuristic} by making it seem omnipresent \cite{lewandowsky13schwarz}. 
Furthermore, we have seen that it can be difficult for sighted users to understand and interpret information in a voice-only setting compared to individuals with visual impairments~\cite{brag2018listen}.
Another example is a sighted user might only be able to process a few initial sentences and then conclude heavily relying on this information, magnifying the \textit{Position Effect}. 
As a result, providing voice-only responses needs to come with a careful consideration of the potential cognitive biases. 
While the influence and mitigation of cognitive biases in web searches have been extensively researched, there is still a significant gap in our understanding of the role of such biases in SCS, i.e., a voice-only setting.

\begin{figure}[tp]
    \centering
    \includegraphics[width=0.7\linewidth]{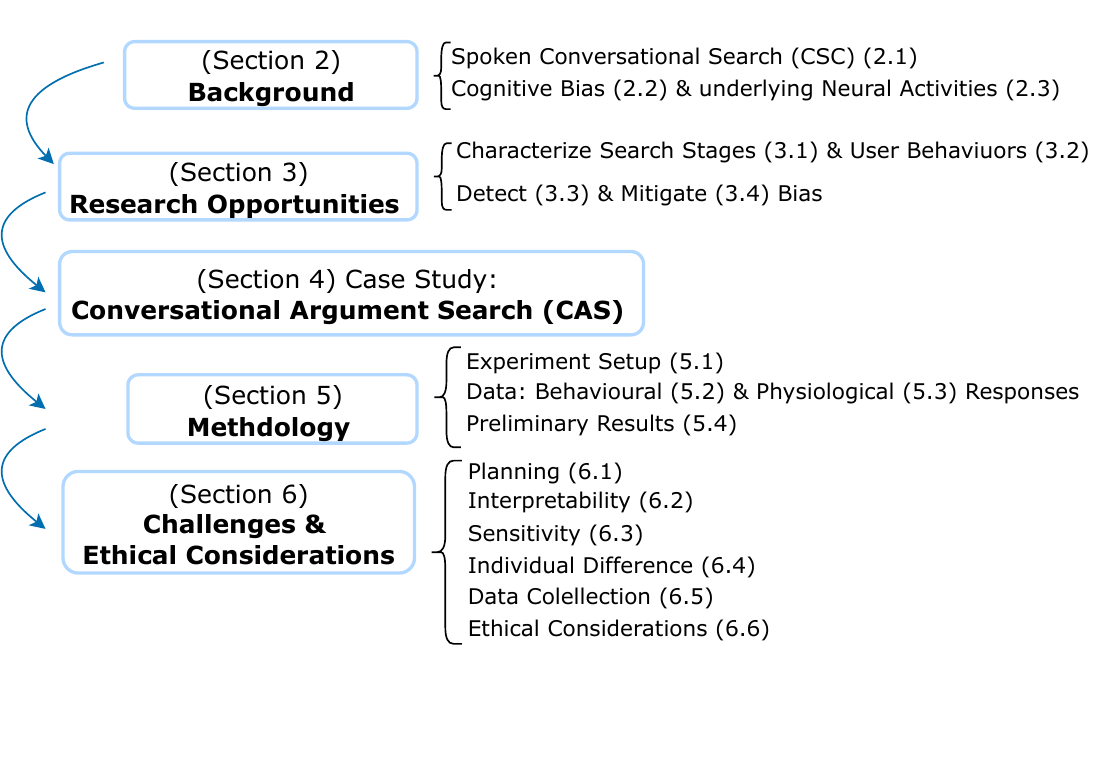}
    \caption{Structure Diagram of this paper.}
    \label{fig:structure}
\end{figure}

An essential challenge lies in the experiment.
Screen-based web search benefits from well-defined tools, such as eye-tracking~\cite{buscher2012attentive, cole2014task, zillich2021selective, harris2019eyetracking} and click-through logs~\cite{suzuki2021characterizing, lewandowsky13schwarz, draws2021not}, as well as standard protocols to visualize and study behaviors, to determine the influence of cognitive biases \cite{suzuki2021characterizing}. However, such methodologies are not established for SCS, which calls for instruments, methodology, and protocols that go beyond the visual paradigm. The remainder of this paper is organized as shown in Figure~\ref{fig:structure}. 

This paper presents new methodologies, novel applications of tools (e.g., wearables), demonstrated through the following three contributions:
\begin{enumerate}
    \item [(i)] We discuss the tools that have been employed to study user behavior and cognitive biases during web search, and consider their use in the context of SCS. We then identify research opportunities that lie ahead in the context of exploring cognitive biases in SCS.
    \item [(ii)] We propose approaches for researchers and practitioners to instantiate experimental designs, from the setup format to the type of user signals. Using the proposed settings, we discuss preliminary results from a laboratory study demonstrating the potential of using wearable devices such as an electroencephalogram (EEG) as a voice channel equivalent to eye-tracking in web search.
    \item [(iii)] We outline the challenges and offer suggestions for adopting our approach and ethical considerations, to achieve accurate and representative results from multimodal signals.
\end{enumerate}

\section{Background}
\label{sec:relatedwork}

\subsection{Spoken Conversational Search}
\label{subsec:Conversational Information Seeking}

Conversational information seeking (CIS), the process of obtaining information through conversations (text, audio/voice, or multi-modal), is a fast-developing research area~\cite{radlinski2017theoretical, zamani2023conversational}. CIS supports users to search for information through natural language with a search system. A CIS system enables users to ask questions, refine their questions, ask follow-up questions, or provide relevant feedback in a natural manner. 
Such systems could either be single-turn or multi-turn. 
In contrast to a single-turn, a multi-turn setting typically maintains the conversational context (e.g., co-reference resolution)\footnote{By ``context'', we mean the information exchanged during the conversation necessary to interpret the users' response, e.g., the history, preferences, and so on.} in a back-and-forth information exchange with the user~\cite{zamani2023conversational}. 
Some advantages of multi-turn CIS include alleviating the cognitive burden on the user by breaking down the information, assisting with information need formulation, or providing highly personalized information for a given context~\cite{trippas2020towards}.
While context management may be relatively trivial for a CIS system, users also have to perform context management subconsciously. This would require significant cognitive effort from the user, particularly when the conversation gets longer, and the task gets more complex.

This paper focuses on SCS, a type of CIS, where communication between the user and system is entirely mediated verbally through audio~\cite{trippas2018informing}. Visual CIS interfaces often use screen-based cues like boldfacing important sections of text~\cite{chuklin2019using}, or attributing sources
within the responses of text-based CIS~\cite{lewis200rag}, including large language model (LLM) based conversational agents~\cite{shah2022situating,liao2023proactive,benedict2023generative}.
These cues help users effortlessly identify the required information. However, a voice channel like that of SCS poses new challenges. Specifically, because it is linear in nature, users may struggle to keep up with the information presented. Additionally, in SCS, the amount of information that can be conveyed is limited by users' cognitive capacity. Recent research by~\citet{chuklin2019using} also found that the features of the audio used by the system to communicate with users can affect their understanding.

\subsection{Cognitive Biases in Information Seeking}
\label{subsec:Cognitive Biases}

Cognitive biases can also be referred to as a ``pattern of deviation''~\cite{belief2013white, kahneman2021noise} from norm or rationality in thinking and reasoning, which influences the decisions made. It is based on a well-known assumption that humans have limited cognitive capacity~\cite{SWELLER201137}, so they tend to favor mental shortcuts (e.g., other judgments like system ranking, or crowd opinions) 
~\cite{biasreview2021,soprano2024cognitive}.\footnote{See \citet{biasreview2021} for a more comprehensive description of various cognitive biases in search.}
In SCS and Information seeking in general, users search for and access information that is perceived as trustworthy. Further in this process, users build mental models around the links between different pieces of information~\cite{lewandowsky13schwarz}, and factors like cognitive bias would influence the entire information consumption process~\cite{biasreview2021}. 
These biases influenced fair information receiving, i.e., having balanced views, equal trust, and cautious evaluation of any information/sources when users search for decision-making. During this process, the information cherry-picking will likely be affected by the order (rank) (\textit{Order Effect}), the balance or imbalance in exposure (\textit{Exposure Effect}), a prior judgment (\textit{Confirmation Bias}), the first piece of information (\textit{Anchoring Bias}) or \textit{Misinformation}~\cite{lewandowsky13schwarz, novin2017making, case2005avoiding,soprano2024cognitive}.

Examples of the negative influence of biases include citizens supporting a partisan without thoroughly evaluating their policies, reinforcing stereotypes against minority groups, and spreading misinformation leading to potential disturbances \cite{selective2013Review, biasreview2021, lewandowsky13schwarz}.
It is imperative that people participate with an open mind and seek out reliable sources of information.
From the user experience perspective, cognitive bias can expedite the process of absorbing information.
A user might initiate bias to quickly and effectively move through the overwhelming information. 
Thus, it is necessary to investigate the impact of biases and provide accurate in situ information when building a personalized informatics system. 

The common approach to measuring cognitive bias in web search is web-logging~\cite{epstein2015search, suzuki2021characterizing}. By analyzing the sentiments on each search query or result, dwell time on search engine results page (SERP) or webpages, search session time, first-clicks and repeat-clicks, existing works have investigated 
\textit{Exposure Effect}~\cite{draws2021not}, 
\textit{Search Engine Manipulation Effect}~\cite{epstein2015search}, and 
\textit{Confirmation Bias}~\cite{suzuki2021characterizing}. 
However, web logging has shown its limitation in less granular data. For example, a study by~\citet{suzuki2021characterizing} did not observe signs of \textit{Confirmation Bias} with web logging, because participants mainly used neutral-attitude queries and similar time on reading each stance. However, comparing dwell time alone cannot accurately represent bias since users may spend longer on opposing opinions to offer criticism~\cite{selective2013Review}. In that sense, web logging can not provide details about what has actually happened. Our review of research on cognitive biases within information access, summarized in Table~\ref{tab:literature}, highlights a lack of exploration concerning cognitive biases in voice-based systems. This existing gap in research serves as motivation for us to investigate this issue further and provoke further conversations within the IR community. Recognizing the limitations of traditional tools for exploring cognitive biases in traditional web search for voice-based systems we emphasize the importance of leveraging neural activity. In the following section, we delve into the nature of these neural signals and advocate for their utilization in studying cognitive biases in SCS.

\subsection{Neural Activities for Cognitive Bias}
\label{subsec:Neuro-activities of Cognitive Bias}

Neural activities have attracted attention when studying information seeking. The human brain is divided into several regions in charge of different functionalities. For example, the frontal lobe is associated with executive functions, including decision-making, motivation, and focus. The temporal lobe is associated with auditory and language processing~\cite{MARTINEZMALDONADO202046}. 
The brain processes information received through the senses, which are then converted into electrical signals. These signals travel along neurons and are sequentially processed in various brain regions. Each region's activity generates electrical and chemical changes that produce detectable signals. These signals can be measured with specialized equipment such as as electroencephalogram (discussed in Section~\ref{subsec:multi-modal sensing}) and functional magnetic resonance imaging (fMRI). These signals provide a window into the flow and processing of information within the brain~\cite{luck_kappenman_2016, randall2014putting}. 
By investigating these neural activities, \citet{jimenez2018using} measure the workload change in web browsing activities.
\citet{moshfeghi2013effective} discriminate different search intentions regarding physiological arousal and emotions captured from multi-modal sensing. 
Further, \citet{moshfeghi2019towards} identify the brain regions and the neural activities involved when participants realize whether they need to search for information. 
\citet{ye2022towards} investigated the brain reaction when people identify keywords relevant to the information they seek.

Regarding audio processing, listening effort refers to the cognitive resources people spend on listening~\cite{picou2011visual, francis2020listening}. The audio information is first stored as a ``buffer'' in working memory, then processed for comprehension, and then potentially stored in long-term memory~\cite{regev2019propagation, picou2011visual}. During this process,  information that is discrepant with the current mental model or perceived as non-relevant will not be processed~\cite{regev2019propagation}, or requires more effort~\cite{richter2017comprehension} to interpret the information further.
Compared to most individuals with a visual impairment, sighted users have a reduced ability to understand and interpret information via listening~\cite{brag2018listen}. They spend more cognitive effort on audio-only scenarios (e.g., less information is recalled~\cite{schneiders2020remains}), making them more vulnerable in voice interactions. This increased effort spent on understanding and interpreting information can hinder their ability to allocate enough effort toward reasoning and critical thinking, which are crucial for mitigating cognitive bias~\cite{biasreview2021}.

Regarding cognitive bias, we can understand whether the user has encountered bias by understanding what cognitive activities are involved in each context, e.g., whether the information only reaches the language region, or gets passed to memorizing and memory-retrieving.
\citet{moravec2018fake} found that participants spent more cognitive activities to judge the trustfulness of the news headlines that supported their beliefs and allocated more attention. 
\citet{jiang2020believe} revealed that 
the receiver's attentional resources are quickly allocated to decode utterances if the speaker has a confident voice. In contrast, other cognitive resources are required to decode voice cues and utterances if the speaker has a doubtful voice. Furthermore, the study discovered that initial judgments are made at the beginning of an utterance and can potentially influence the final decision. This implies the development of a bias during audio processing and highlights the potential of investigating cognitive bias in voice search.
Given this interpretation, \citet{randall2014putting} investigated the cognitive processes that contribute to the development of cognitive bias in decision-making. As previously discussed, the brain regions responsible for language processing activate first, followed by comprehension and assessment of working memory. If the information is deemed irrelevant or dissident, it will be discarded and not proceed to the next level or other regions. Their experimental findings have supported this hypothesized process.

With advances in wearable devices, physiological sensing has been adopted in detecting cognitive bias in web searches, grounded with assumptions such as 
cognitive load theory~\cite{SWELLER201137}, 
orienting responses~\cite{sokolov2002orienting}, 
cognitive dissonance (dissonance arousal)~\cite{elkin1986physiological, ploger2021psychophysiological}, 
and dual-thinking system theory~\cite{kahneman2017thinking}.
Furthermore, \citet{boon2023bias} and~\citet{ploger2021psychophysiological} found increased skin conductance levels on the presentation of a dissenting stimulus, which suggests dissonance arousal. However, in a different context, decision-making in a team discussion, \citet{randall2014putting} found increased skin conductance, indicating greater emotional arousal, on the supporting stimulus, suggesting that the participants become more activated to think about the supporting information.
Apart from skin conductance, facial Electromyography (EMG)~\cite{randall2014putting} and heart rate variability (HRV)~\cite{ploger2021psychophysiological} are also employed in experiments. Multi-modal data are discussed later in Section~\ref{subsec:multi-modal sensing}.

\section{Research Opportunities}
\label{sec:Challenges and Opportunities}

Characterizing cognitive biases in SCS presents research opportunities, including novel ways to characterize search stages, user behavior, and new cognitive bias detection or mitigation~strategies.

\subsection{How to Characterize Cognitive Bias at the Different Stages of the SCS Process?
}

Cognitive bias may occur at each stage of a visual-based search process~\cite{biasreview2021}, i.e., querying, consuming the search results, and judging relevance and satisfaction. 
It has been suggested that search stages or actions~\cite{maxwell2016agents} (i.e., query application and reformulation, snippet scanning and assessment, snippet clicking, document reading, document assessment, and session stopping) and user behaviors differ for screen and audio-only channels~\cite{trippas2020towards}.

Similarly, cognitive biases manifest differently in these search stages for screen and audio-only channels~\cite{khofman2023exploring}.
For instance, with screens, the users can review and easily refer back to their query, which is inherently more difficult with voice queries~\cite{sa2020examining}. In SCS, the queries are more likely to be posed in natural language~\cite{gud2018the, crestani2006written}, and the syntactical arrangement query words may indicate the user's intent~\cite{smirova2020word} and perhaps even reveal any underlying biases. Another instance is when one might ask, ``Why is renewable energy inefficient?'' rather than ``What are the efficiencies of renewable energy?''. This could suggest that a user's preconceived beliefs about a topic, influence the querying stage, resulting in biased search results. Moreover, the query stage might further be influenced by a user's false memories (i.e., attributes the user misremembered about a searched item), presenting an additional challenge for search systems~\cite{kiesel2019falsememories}. \citet{kiesel2019falsememories}~noted that users may not easily accept misremembering something. To this end, we highlight the significance of exploring cognitive processes at different stages of SCS interaction, such as detecting false memories at the query stage. 

\subsection{What Is 
the Role of Clarifying Questions in SCS? How Is It Related to Cognitive Bias?}
\begin{figure}[tp]
    \centering
    \includegraphics[width=0.6\linewidth]{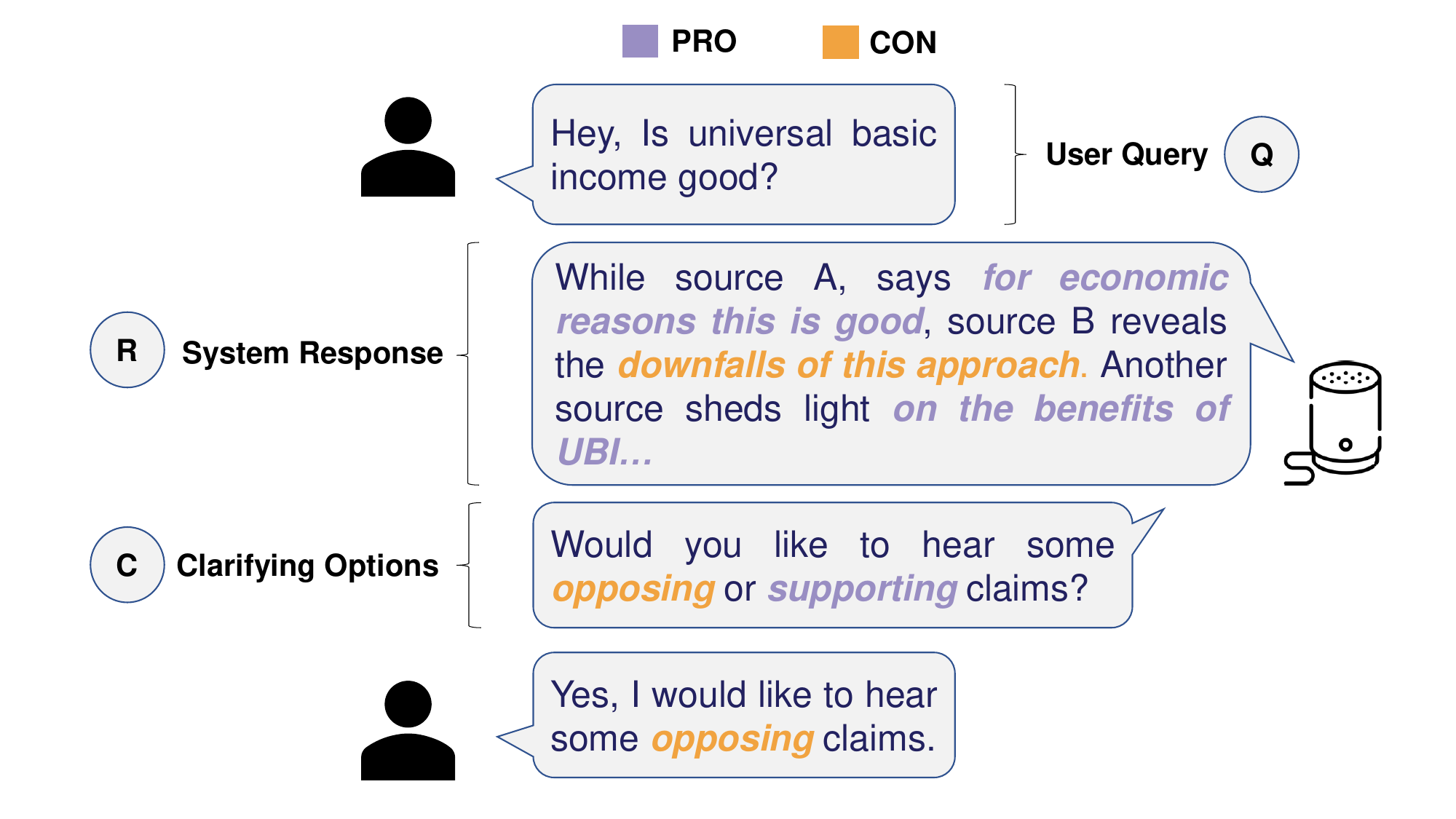}
    \setlength{\abovecaptionskip}{5pt}
    \caption{Example interaction between a user and a SCAS system involving two perspectives, i.e., supporting (\pro) and opposing (\con).}
    \label{fig:cas}
\end{figure}

In CIS, the inherently dialogic nature of interactions means that reformulating queries and asking clarifying questions become more critical and ideally occur more frequently, supporting conversational actions~\cite{zamani2023conversational, alian2022conver, trippas2015resultspresentation}.
Users often refine their queries by referring back to previous responses, seeking to narrow down or expand upon their initial inquiry~\cite{zamani2023conversational} as seen in the~Figure~\ref{fig:cas}C.
Cognitive biases may influence this iterative process. For example, if users receive information that aligns with their pre-existing beliefs, users may accept the response without further questioning. Conversely, if the information opposes their beliefs, they may reformulate their query to obtain results that align more closely with their expectations. This means that considering a user's reformulation/clarifying questions can help to detect potential bias. 
This implies the presentation of clarifying options becomes as important as presenting users with relevant responses to satisfy the user's information need. In the context of SCS, different ways of providing clarification options influence user satisfaction~\cite{kiesel2018voiceclarifying}, however, the arrangement and format of the options presented to the user may contribute to the reinforcement of \textit{confirmation bias}, which is a research challenge that has not been fully explored.

\subsection{Can Voice Modulation Be Used to Characterize Cognitive Bias?}
While eye-tracking data is not feasible in voice interactions, attributes related to the audio signal (e.g., pitch and speed), from both the system and user, can reveal rich information regarding motivations, emotion, and personal traits~\cite{leongomez2021voice}. For example, \citet{jiang2020believe} indicated that the perceived \textit{information believability} is impacted by whether the voice sounds confident or doubtful. Moreover, \citet{GOODMAN2023103864} found that the agents with female voices received higher trust than male voices. And as the pitch increases, participants are less likely to make decisions with the information provided by the agent. These examples suggest that the system's voice modulation influences how people perceive information. Currently, we do not know how the system's voice modulations may affect a user's beliefs or reinforce certain biases such as \textit{confirmation bias}, making it an open research problem.

A potential solution is slowing down the system when discussing controversial opinions, giving the user sufficient time to absorb and consider the different aspects.
A further important direction is investigating how cognitive biases are related to the voice modulation of a user. For example, a user speaking in a skeptical tone and a higher pitch to ask the system, ``Is climate change REALLY [accentuate] happening?'', may indicate that the user has a certain degree of \textit{confirmation bias} towards their pre-existing belief that climate change is not a real problem.

\subsection{How to Leverage Content Manipulation to Mitigate Harms of Cognitive Bias?}
Cognitive bias need not solely have a negative effect~\cite{metzger2013credibility}. While it is true that biases can lead to skewed perceptions and decisions, they can also serve as tools for balancing perspectives~\cite{Kiesel2021conv}. For example, \textit{Availability Bias} refers to placing greater importance on readily available or easily recalled information. One way to counteract this is by intentionally presenting less readily available information first. However, we need to consider whether this solution may create new issues related to group fairness or the spread of misinformation. 
Recognizing and understanding the impact of cognitive bias helps address potential pitfalls and leverage its potential to create an effective and user-friendly search experience.
In the context of SCS, auditory icons (i.e., earcons) may be used to design mitigation strategies for unintended cognitive bias. A recent study by~\citet{gohsen:2023a} explored the effects of different audio interventions (nudges) to offer guidance to seekers in spoken conversations. Through an elaborate crowdsourced experiment, they identified that the most effective nudging technique was explicitly suggesting questions. Additionally, \citet{gohsen:2023a} used the NASA-TLX to measure the seeker's mental load. The results showed that the mental load for all the nudging techniques was approximately uniformly distributed, and consequently, the authors raise the question of whether NASA-TLX is an appropriate metric. This may indicate that self-reported measures are not always reliable, and calls for capturing more fine-grained information using wearable technology (e.g., EEG).

\section{Case Study: Argument Search}
\label{subsec:Argument Search}


Building on the research opportunities we have identified, we introduce a specific use case of SCS called Spoken Conversational \textit{Argumentative} Search (SCAS). This use case will provide a clear idea of the data, topics, and other metadata involved in designing experiments within the framework of our proposed approach, which we then discuss in detail in Section \ref{sec:Experimental Framework for Understanding Cognitive Biases in CIS}.

SCAS systems respond to a user's spoken query on controversial topics with multiple argument stances or viewpoints (i.e., \pro and \con). During the interaction, biased exposure of these stances can lead to misrepresentation, and potentially lead to broader negative effects, including on society \cite{draws2021not,yunkaporta2023right}. The sources of biases can arise from the quality of the data (e.g., correctness or completeness), the underlying algorithms~\cite{pathiyan2021argfairness}, and how the information is presented to the user. 
In the era of information overload, a biased system can lead to polarized societies, reinforce stereotypes (e.g., stereotypical gender biases~\cite{bigdeli2023understanding}) and even influence public opinion on critical matters. This type of unconscious bias may lead to discrimination against particular groups, such as people with a visual impairment. While people with a cognitive or visual impairment may directly benefit from SCAS, its utility also extends to sighted users. For instance, podcasts are attracting rapidly increasing numbers of listeners, and there is evidence that podcast consumption impacts political leaning~\cite{chadha2012listening}. When engaged in activities that preclude reading, such as driving, cooking, or even exercising, users can rely on SCAS to provide them with balanced arguments on topics of interest. Moreover, future learning environments imply the potential of SCAS in educational settings. In a learning context, it is crucial to teach students with unbiased information and diverse perspectives \cite{morgan2009picture}. An example is a user asking the SCAS system ``is universal basic income good for society?''. If the existing system only provides one side (for instance, \pro) of the issue, the user tends to be blind-sided by not having any information about other perspectives~\cite{gao2020towards}. The imbalanced exposure of perspectives is an important open challenge for SCAS \cite{pathiyan2022fairness}.

\raggedbottom

\subsection{Data}
 For our specific use case, designing experiments requires argumentative topics  (e.g., ``Is universal basic income good for the society?'') and documents/passages supporting (\pro) and opposing (\con) the topics. 

A recent crowdsourced study conducted by \citet{draws2021not} involved 100 participants rating their opinions on 18 argumentative topics from the ProCon.org debate portal\footnote{\url{https://www.procon.org/debate-topics/} [Accessed: 9 Feb 2024]} using a Likert scale. The study identified that 5 of these topics had only mild pre-existing viewpoints among the participants. We suggest that incorporating such topics into future cognitive bias experiments is highly important to help avoid heavily polarized topics, and therefore facilitate detecting the effects of certain cognitive biases more effectively. Since the existing data only includes 280 search results, which may not be sufficient if the participant in an experiment is required to engage with the setup in longer conversations.
Therefore, we propose extending the collection by using the args.me corpus \cite{ajjour:2019a}. The args.me corpus was created for studying argument retrieval and was utilized for the shared tasks at \touche @ CLEF 2022 \cite{bondarenko2022overview}.\footnote{The args.me corpus \cite{ajjour:2019a} consists of arguments that are associated with a stance (i.e., \pro or \con). It contains 387,740 arguments crawled from four debate portals (debatewise.org, idebate.org, debatepedia.org, and debate.org) and 48 arguments from Canadian parliament discussions.} Meanwhile, the \touche @ CLEF 2023, focused on retrieving argumentative web documents from the web crawl corpus ClueWeb22 \cite{bondarenko2023overview}, making the \touche @ CLEF 2023 corpus unsuitable for our proposed approach. From an IR perspective, argument retrieval is the task of retrieving relevant supporting (\pro) and attacking (\con) justifications (premises) for a given query (claim) \cite{pathiyan2021argfairness}. Furthermore, the collection provides detailed information on the subtopics discussed within arguments, in addition to stance. For instance, an argument could be justified using multiple sub-topics (i.e., Tax, Capitalism, Healthcare, Poverty, and so on). This means it is crucial to control participants' exposure to subtopics, as much as the stance, in future experiments to account for unknown effects.

\section{Methodology}
\label{sec:Experimental Framework for Understanding Cognitive Biases in CIS}

This section introduces a general experimental approach that researchers and practitioners can use to investigate cognitive biases in SCS, including experimental design, and the collecting of behavioral and physiological data. As an example of our approach and to demonstrate the potential of physiological data, we also present preliminary results from an information-seeking experiment we have conducted.

\subsection{Experiment Setup}
To accommodate various needs of research questions and their associated experiments,  
including feasibility, scalability, research method (qualitative, quantitative, mixed), 
we discuss several potential set-ups along with their advantages and disadvantages.

\subsubsection{Lab Study}
\begin{figure}[tp]
        \centering
        \includegraphics[width=0.4\textwidth]{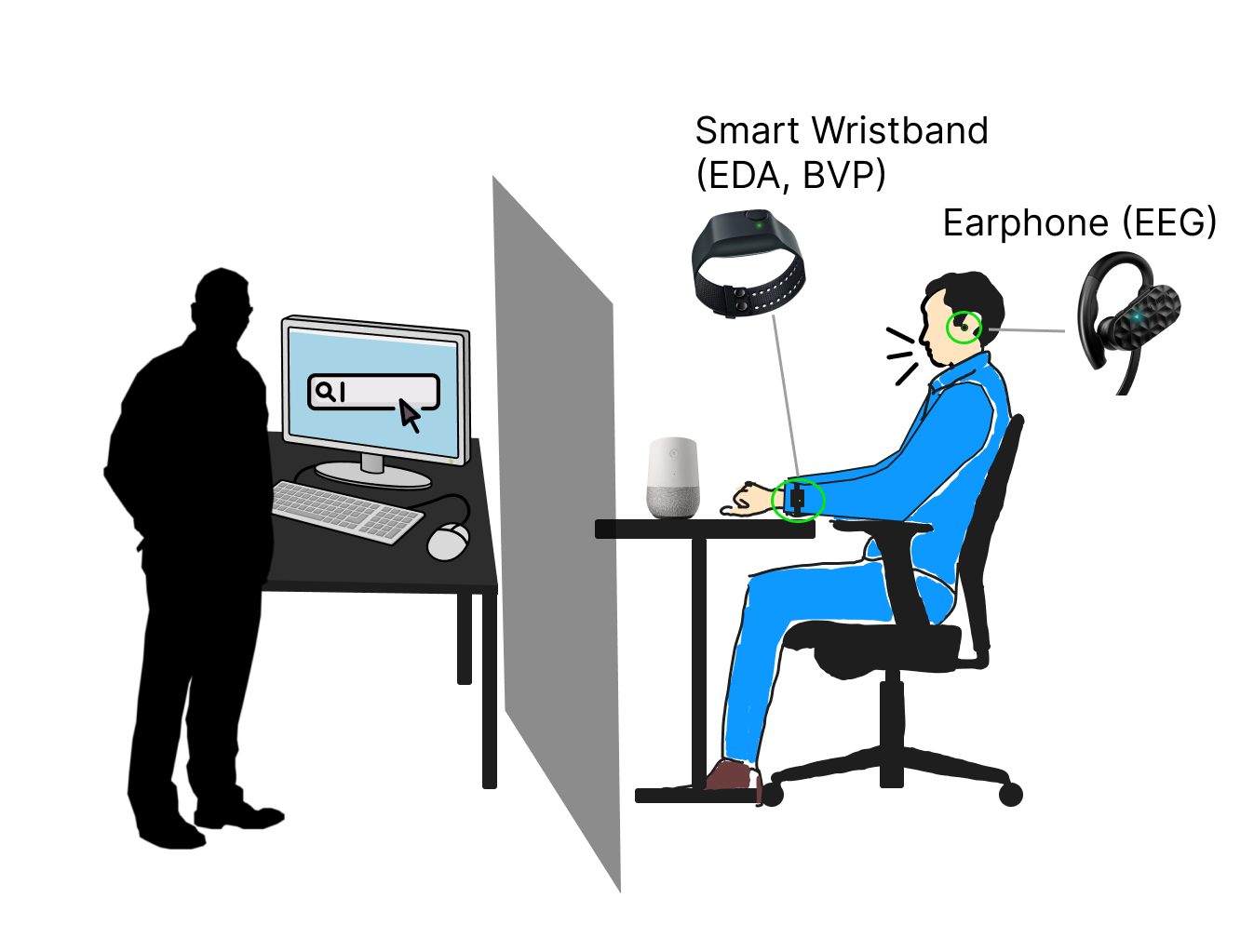}
        \caption{Wizard of Oz (WOZ) Set-up where the right side depicts a participant wearing sensors while interacting with the SCS system. The left side depicts the intermediary placed separately from the participant to simulate an information access system that provides a response based on the user queries.}
        \label{fig:woz}
    \end{figure}

Multiple qualitative methodologies have been proposed to examine how users behave during a conversation. For instance, ~\citet{trippas2020towards} outline a lab-based data collection and a qualitative analysis framework for SCS.
For a lab study, one popular technique is the Wizard of Oz (WOZ)~\cite{dubiel2018investigating, vtyurina2017exploring, trippas2017people} study. As demonstrated in Figure~\ref{fig:woz}, users interact with an information access system through an intermediary. The intermediary receives the utterances from a user and performs the search task like a conversational system would. This WOZ setup helps to overcome the technical difficulties of transcribing a user's voice-based query to text. 
The controlled setup in a lab study also enables capturing the physiological responses of users~\cite{ji2023towards, Boonprakong2023exploringindicators}.

\raggedbottom

\subsubsection{Field Study}
The participants can be given pre-configured voice agents (such as the experiment by \citet{wei2022esm}) and wearable devices, to take them home. If the devices are easy to wear, the setting can be extended to conduct longitudinal studies. Accurate remote data collection mechanisms and time-syncing capabilities are needed to run successful field studies. 
While SCS field studies are more ecologically valid when compared to lab studies, one potential challenge is dealing with noisy environments and real-world data. Unlike lab studies, various unobserved factors in the real environment can obfuscate the effects of interest in the data, making the analysis more complex.

\subsubsection{Crowdsourcing Study}

Lab studies such as the WOZ technique are resource-intensive, as they require participants to be physically present, and typically require further supporting staff. This may restrict scalability and prevent researchers from expanding the study to a large number of participants. 
Given the rising interest in studying SCS systems and associated user behavior within the research community, the issue of scalability is important to consider. 
With the introduction of sensor-embedded consumer devices (e.g., Apple Airpods embedded with EEG~\cite{azemi2023biosignal}), sensors are becoming more accessible for daily use. The future of studying the SCS system using a crowdsourced setup is likely to become increasingly viable, and it appears we are already heading in that direction. Earlier works, such as the study by \citet{Madiha2020alwayslisteningva}, utilized crowdsourcing platforms like Prolific to examine the efficacy and necessity of always-on voice assistants by placing users in a hypothetical environment. Additionally, recent work by \citet{Hettiachchi2020} also proposed the need for a voice-based crowdsourcing platform that runs on a digital voice assistant. Unlike lab or field studies, crowdsourcing lacks live intermediary for controlled responses. With the recent advances in Large Language Model (LLM) based applications such as Retrieval Augmented Generation (RAG) \cite{lewis2020retrieval}, we see the potential to control the study as required for the research question being explored as demonstrated earlier \cite{julius2024jaybot,pathiyan2024walert}. Such work highlights the potential of conducting crowdsourcing experiments for SCS. Additionally, such a RAG-based SCS system can also be helpful in laboratory and field-based studies by assisting intermediaries in processing a retrieved list of documents and generating a final response. One of the challenges in a crowdsourced setting is to maintain or increase the ecological validity, as previous studies have shown the presence of a physical entity (i.e., the smart speaker) can impact how users interact with the system and perceive the information~\cite{sunok2020usermentalmodel}.

\begin{table}[tp]
\caption{A breakdown of the measures studied in the literature based on the type of data used (Behavioral and Physiological), and mode of user interaction (screen-based vs voice). The bold text in the table denotes the studies focusing on cognitive biases and highlights the lack of research on cognitive biases in voice search.}
\label{tab:literature}
\resizebox{.95\textwidth}{!}{
\begin{tabular}{cp{7cm}llll}
\toprule
 \multicolumn{2}{c}{\multirow{2}{*}{Data Type}} & \multicolumn{2}{c}{\textbf{Screen-based}}    & \multicolumn{2}{c}{\textbf{Voice}}     \\ \cmidrule(lr){3-4}\cmidrule(lr){5-6}
 &    & \multicolumn{1}{l}{Construct} & \multicolumn{1}{l}{Related Work}     & \multicolumn{1}{l}{Construct} & \multicolumn{1}{l}{Related Work}    \\ \midrule
\multirow{7}{*}{\begin{turn}{90} \textbf{Behavioral}\end{turn}} &  Web-logging (e.g., dwell time, clicks)  & \textbf{Cognitive Bias} & \cite{suzuki2021characterizing, lewandowsky13schwarz, draws2021not}           & \multicolumn{2}{c}{--}    \\ \cmidrule(l){2-6}
                               & Transcripts \& Voice Modulation (e.g., pitch, speed)  & \multicolumn{2}{c}{--}      & Perceived Trust & \cite{GOODMAN2023103864,lindgren2023intimacy}       \\  \cmidrule(l){2-6}

 &  \multirow{2}{*}{\begin{tabular}[l]{@{}l@{}}Task Performance (e.g., sentiments of \\ query/utterance, recall rate)\end{tabular}}   &      \textbf{Cognitive Bias} & \cite{epstein2015search}   & Listening Effort  & \cite{brag2018listen, schneiders2020remains, kiesel2018voiceclarifying, picou2011visual, karat1999patterns}  \\ 
  &    &      Search Experience & \cite{melumad2023vocalizing, Setlur2022analyticalchatbot}   &   Search Experience & \cite{ Setlur2022analyticalchatbot, karat1999patterns}  \\ 
 \cmidrule(l){2-6}
                                & Motion, Facial Expression, Gaze                 &         \multicolumn{2}{c}{--}      & Engagement   & \cite{ooko2011estimating, gazehumanagent2010, ooko2011estimating}              \\  
                                    
                               \midrule
             
                      \multirow{9}{*}{\begin{turn}{90} \textbf{Physiological}\end{turn}}  &          \multirow{3}{*}{Brain Signals (e.g., EEG)}         & Cognitive Workload      & \cite{jimenez2018using, mohamed2018characterizing}     & Perceived Trust      & \cite{jiang2020believe}        \\  
                             &  & Search Experience & \cite{moshfeghi2013effective, Dominika2022information, ye2022towards, gwizdka2017temporal, marco2015relevance}                &    &      \\ 
                               & & \textbf{Cognitive Bias}   & \cite{moravec2018fake, randall2014putting, bago2018fast, yu2023people}               &          &           \\  \cmidrule{2-6}
                                & Peripheral Sensing (e.g., EDA, PPG)  & \textbf{Cognitive Bias}     & \cite{randall2014putting, boon2023bias, ploger2021psychophysiological}         &   \multicolumn{2}{c}{--}                \\                
                              \cmidrule(l){2-6}
                              &  \multirow{3}{*}{Pupillary Responses}  & \multirow{3}{*}{Selective Attention}     &  \multirow{3}{*}{\cite{regev2019propagation, gwizdka2017temporal}}            &     Selective Attention    & \cite{regev2019propagation}       \\   
                              &  &  &     & Distraction     & \cite{marois2019auditory}  \\         
                             &  & &   & Listening Effort   & \cite{picou2011visual}       \\  
                             \bottomrule
\end{tabular}
}
\end{table}

\subsection{Behavioral Responses}
\label{subsec:tracking_behavioural}

Data collected during SCS experiments can be grouped in Behavioral Responses (discussed in this section), and Physiological Responses (discussed in the following section). The groupings, data types, and measures are summarized in Table~\ref{tab:literature}.

Compared to screen-based interaction, analyzing audio-only interactions collected in SCS is less straightforward.
SCS lacks easily observable indicators of comprehension or focus, i.e., the listening effort (refer to Section~\ref{subsec:Neuro-activities of Cognitive Bias}). Moreover, conversational interactions are less structured, making it more challenging to identify and measure specific biases. Although web-logging data are unavailable for SCS, other types of behavioral data can provide a wealth of information. 

\subsubsection{Transcripts \& Voice Modulation}
On screen-based search, task performance \cite{epstein2015search} or web-logging data, e.g., clicked search results \cite{suzuki2021characterizing, lewandowsky13schwarz, draws2021not}  
have been used to detect bias. 
In voice-based search, \emph{utterances} serve as queries \cite{sa2020examining}. Voice modulation is another attribute that screen-based search does not have, as discussed in Section~\ref{sec:Challenges and Opportunities}.

\citet{wei2022esm} observed that users would raise their voices, repeat the questions, reformulate the queries, or change the pronunciation when dealing with voice assistance errors. 
A \textit{system error} occurs when the system fails to provide users with the desired results, regardless of whether the error is caused by an incorrect response or lack of a response. If we see bias as a \textit{system error} where the system does not give users the answer they want. This can occur when the answers do not align with users' pre-existing beliefs or when the answers are too long, and users don't have the patience to listen to everything. In some cases, users may not be able to recognize that issues are due to cognitive bias in the design process. Instead, they may report these problems as a 'system error', especially for users with lower technology experience. Anyhow, in such cases, users may attempt to bypass the system to obtain their desired results. For example, they may ask for a shorter version of the answer, which can potentially produce bias.
In short, transcripts and voice modulations provide rich information to understand the interactions and differentiate biases from system errors in SCS.

\subsubsection{Task Performance}
When measuring listening effort or speech intelligibility, researchers usually use recall/remember tasks, including individual word recognition, sentence comprehension, and sentence recognition \cite{brag2018listen}.
By assessing the accuracy of information the participants recall or recognize, we can capture which piece of information received more attention, and provide some hints on the cognitive processing related to language \cite{regev2019propagation}. 
These tasks have the potential to be suitable for investigating bias.
Under similar conditions (such as complexity and speech intelligence), it is assumed that people are more likely to accept and comprehend certain information when encountering bias. As a result, they tend to remember and recall that information more accurately \cite{bago2018fast}. 
Specifically, biased information receives more attention and requires less cognitive load to process. Information that is consistent with participants' existing beliefs tends to attract more interest and engagement and is easier to remember (\textit{Confirmation Bias}), whereas information that is inconsistent is more difficult to process and requires more cognitive effort.
However, recall tasks may lack granularity for the detection of cognitive bias.
Additionally, other confound variables may play a role in individual's listening performance~\cite{brag2018listen}, e.g., language proficiency~\cite{kiesel2018voiceclarifying}, and working memory capacity~\cite{federmeier2020examining, picou2011visual}. \citeauthor{sriram2009brief} proposed a new version of the Implicit Association Test~\cite{greenwald1998measuring} -- Brief-IAT~\cite{sriram2009brief}, specifically designed to assess bias. \citet{dingler2022method} proposed a method to appropriately deploy Brief-IAT in crowdsourcing studies. We advocate more such efforts that are necessary to develop tasks specifically crafted to evaluate cognitive bias in the SCS context.

\subsection{Physiological Responses}
\label{subsec:multi-modal sensing}
Cognitive bias can be measured by examining differences in cognitive processes, emotions, and engagements. For example, a user might be more engaged and focused, and experience greater emotionally aroused when the audio reaches the end as opposed to the middle. 
Multi-modal sensing with wearable devices can capture these biological responses and provide a reliable and comprehensive way to `visualize' cognitive bias in SCS, similar to how eye-tracking helps us to understand user interactions with screen-based IR systems.
The data that can potentially be used are described below. 

\subsubsection{Brain Signals}
Electroencephalography (EEG) is a method to collect electrical activity in the brain, and is well-established as a tool for understanding people's brain function -- both cognitive and emotional \cite{luck_kappenman_2016, Bota2019}.
EEGs are often used in scientific studies to understand memory, attention, motivation, emotion, and responses to stimuli (e.g., information or emotional stimuli) \cite{KUMAR20122525}. 
For web search, EEG has shown promising results in detecting relevance judgment at the article level~\cite{gwizdka2017temporal, marco2015relevance}, and word level~\cite{ye2022towards}, as well as identifying information needs in a Q\&A scenario~\cite{Dominika2022information}. \citet{moravec2018fake} observed the increased cognitive activities signifying confirmation bias when judging fake news headlines. 

Two common methods to analyze EEG signals to reveal the activation states of the brain to different extents are Event-Related Potentials (ERP), and Frequency Band Analysis \cite{MORALES2022101067}. ERP is a time-locked analysis that describes the cognitive activity after a particular event's onset \cite{luck_kappenman_2016}. However, ERP is usually used to analyze the signal within a short time window. Studies usually take the grand average of an estimated duration of each word, e.g., 1 second~\cite{ye2022towards, Dominika2022information, eugster2016natural}. However, stimuli for information presentation usually have a longer duration, e.g., 1 minute, and it is difficult to determine a particular onset time for ERP. In such cases, Frequency Band Analysis provides another option. 

\begin{figure}[tp]
        \centering
        \includegraphics[width=0.15\textwidth]{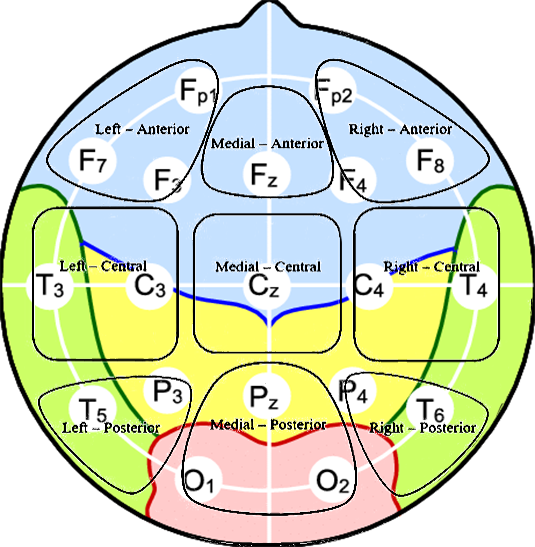}
        \caption{10--20 System for EEG electrode locations and corresponding brain regions~\cite{MARTINEZMALDONADO202046, bos2006eeg}, which could help to characterize the type of brain activities (and corresponding physiology, e.g., cognitive, emotional, and behavioral states).}
        \label{fig:brain}
    \end{figure}

When people engage in different cognitive activities, the brain generates waves of different intensities. There are five commonly categorized frequency bands; in particular, delta (0--4 Hz) is associated with deep sleep, theta (4--8 Hz) is associated with creativity, emotion and memory, alpha (8--13 Hz) is associated with awake or relaxing, and beta (13--30 Hz) is associated with active thinking and problem-solving~\cite{KUMAR20122525}. 
Studies on measuring cognitive bias through EEGs investigated 
the alpha band as an indicator of attention~\cite{randall2014putting, moravec2018fake}, 
the beta band to infer the level of engaging in active thinking~\cite{yu2023people}, as well as 
theta band for the performance of working memory~\cite{mohamed2018characterizing}. 
It is worth noting that the works described above particularly focus on the brain waves generated in the frontal cortex (Figure~\ref{fig:brain}), as they relate to human attention, memory, decoding, and retrieval.

Considering the potential of EEG to offer valuable insights into an individual's brain activity, there has been an increase in the development of these recording products. While traditionally, EEG has been of interest primarily for research purposes, we also see a growing interest in its commercial applications. The headband EEG, e.g., MUSE~\cite{Muse2023}, has been a popular tool for tracking mental wellness and sleep for years. An earbud EEG has recently been released~\cite{Emotiv2023}. In the future, we may see EEG sensors integrated ubiquitously into our earphones~\cite{azemi2023biosignal, BRAIN2022}.
Although it is still unclear how robustly these devices with fewer sensors measure, advancements in technology and data processing might open new doors in SCS. 

\raggedbottom

\subsubsection{Peripheral Sensing}
Commercial wearable sensors can capture a variety of signals that may complement EEG -- e.g., Electrodermal Activity (EDA) / Galvanic Skin Response (GSR), Photoplethysmography (PPG) / Blood Volume Pulse (BVP), Skin Temperature (SKT), captured using a wristband \cite{Bota2019}, e.g., Empatica E4~\cite{Empatica2023};
these are called peripheral signals \cite{ARYA2021100399}. 

EDA captures the variations in the skin's electrical conductance driven by sweat gland activity. This means that emotional responses, especially those triggered by stressful events, often increase perspiration (sweating) and, subsequently, elevated EDA levels \cite{Bota2019, KREIBIG2010394, babaei2021critique}. As such, EDA is commonly used as a sensitive indicator of emotional and cognitive arousal, and the level of engagement. It is observed that EDA decreases when individuals are highly engaged (thus less aroused) in an activity \cite{KREIBIG2010394}. 
PPG is a non-invasive method that uses light to measure blood volume changes. Through PPGs, we can derive heart rate, blood oxygen levels, and other related metrics. PPG can also infer emotion. For example, high arousal or valence (happiness) is associated with a rapid increase in heart rate, which manifests as shorter intervals between PPG peaks \cite{Bota2019, KREIBIG2010394, pham2021heart}. 
Lastly, SKT reflects the balance between the body's heat production and heat loss, and it can be influenced by factors such as emotional states, environmental conditions, and metabolic processes. For example, SKT generally decreases in low valence \cite{KREIBIG2010394}.

\subsubsection{Pupillary Responses}
Pupillary responses have been used to investigate selective attention \cite{regev2019propagation}, auditory distraction \cite{marois2019auditory}, and listening efforts \cite{picou2011visual}. For voice interaction, wearable eye-tracking glasses, e.g., Pupil Labs Neon glasses~\cite{PupilLabs2023}, can provide such a channel. 
However, it's worth considering that pupil data collection is limited by lighting conditions and may only be suitable for lab studies with consistent lighting. Additionally, the cost of using these devices and their impact on participants' physical comfort should be considered. Specifically, it's essential to weigh the benefits of data utility gained from collecting pupillary data against the burden of using additional devices. 
While gaze-tracking may not be as informative in voice interaction as in visual interaction, some devices collect this data by default, resulting in potential data waste.

\subsection{Preliminary Results}

\begin{figure}[tp]
        \centering
       \begin{subfigure}[t]{0.75\textwidth}
            \includegraphics[width=\textwidth]{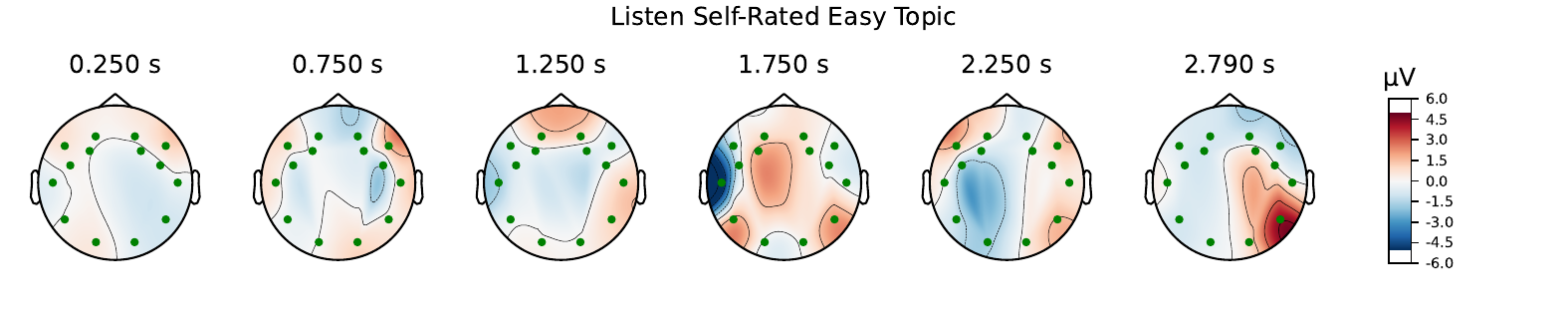}
        \end{subfigure}
        \\
        \begin{subfigure}[b]{0.75\textwidth}
            \includegraphics[width=\textwidth]{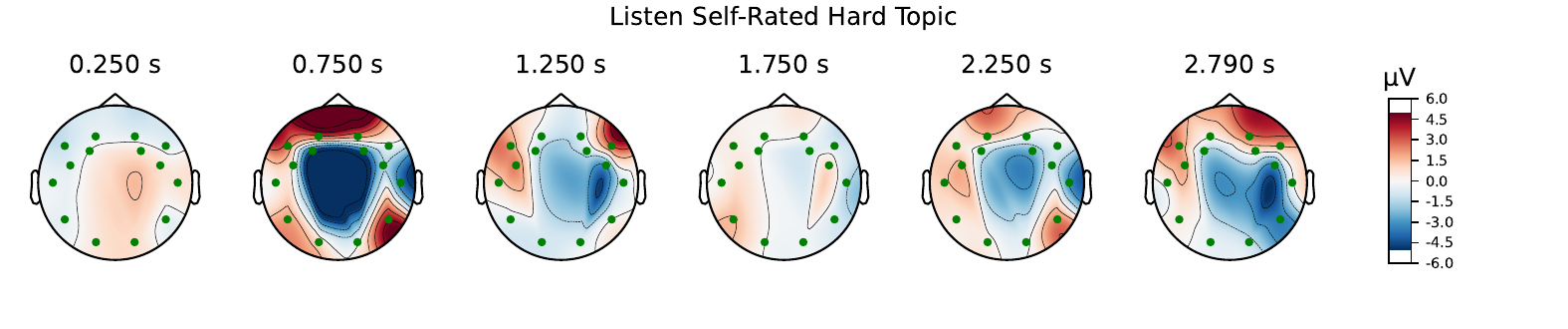}
        \end{subfigure}
    \vspace{-8pt}
    \caption{Preliminary EEG Results ($N=7$) of grand average on listening to Search Results on self-rated $easy$ (Antarctica exploration -- R03.353) and $hard$ topics (Freighter ship registration -- T04.743). Deeper color indicates greater neural activities. Cool colors for negative voltage represent inhibitory activities, i.e., suppressed, blocked, or restricted, while warm colors for positive voltage represent excitatory activities, i.e., promoted, facilitated, or enhanced \cite{luck_kappenman_2016}. The dots represent the placement of 14 electrodes.}
    \label{fig:preresult}
\end{figure}

We conducted a lab user study on information search \cite{ji2024characterizing} 
and collected the data -- including EEG, EDA, BVP, and eye-tracking -- from 7 participants. The apparatus were Emotiv EPOC 14-channel headset, Empatica E4 wristband, and Tobii Fusion Eye-tracker. Although the other data were collected, only EEG and EDA were discussed in this paper for illustration purposes. 

\subsubsection{Procedure} 
The experiment was a simulated information search scenario. The 
participants were given a backstory (context), submitted their search query, and received the audio pre-defined search results (about 1 minute). The participants were required to rate perceived difficulty in understanding each topic and search results from 1 (least) to 5 (most). 
Each participant needed to complete 12 topics in total in random order.

\subsubsection{Data}
Because all stimuli were controlled to be less difficult, we selected results from 2 topics that were self-rated as the most relatively difficult ($\overline{\mu}$ 3.0/5.0) and easy ($\overline{\mu}$ 1.3/5.0).
Although cognitive bias was not the target manipulation in the experiment, the perceived difficulty in understanding the auditory information reveals the change in cognitive efforts in receiving the information. EEG data are cleaned following the procedure described by \citet{eugster2016natural}, then divided into equal-length 3-second epochs (segments) for further analysis. 
EDA data are cleaned and decomposed following the procedure described by \citet{Bota2019}. Subsequently, the EDA and SCL values are aggregated with a 1-second sliding window and then subtracted by the onset value, to present the changes.

\subsubsection{Results}
\paragraph{EEG}
Figure~\ref{fig:preresult} demonstrated a clear difference between the $easy$ and the $hard$ topics. 
Overall, it is suggested that there was less neural activation in the $easy$ topic. Increasing positive voltages were observed at 1.75 seconds in most regions, inferring that more focused attention and engagement. Meanwhile, a pronounced negative voltage in the left temporal lobe might suggest reallocating cognitive resources from processing the auditory information to other brain areas needing more processing power. Then, strong peaks in the right parietal lobe were observed at 2.79 seconds, related to environmental awareness. This can be explained by participants often staring outside of the main task screen when listening to the audio.  
For the $hard$ topic, the heightened neural activation was observed early at 0.75 seconds. Pronounced peaks were observed in the pre-frontal and frontal areas, suggesting participants were engaged in deeper cognitive processing and working memory tasks related to understanding the information. Enhanced activation was also presented in the temporal region, which handles the auditory and language processing; this might indicate increased effort in language comprehension or recalling related knowledge for the $hard$ topic.

\paragraph{EDA}

\begin{figure}[tp]
   \centering
   \begin{subfigure}[]{0.26\textwidth}
        \includegraphics[width=\textwidth]{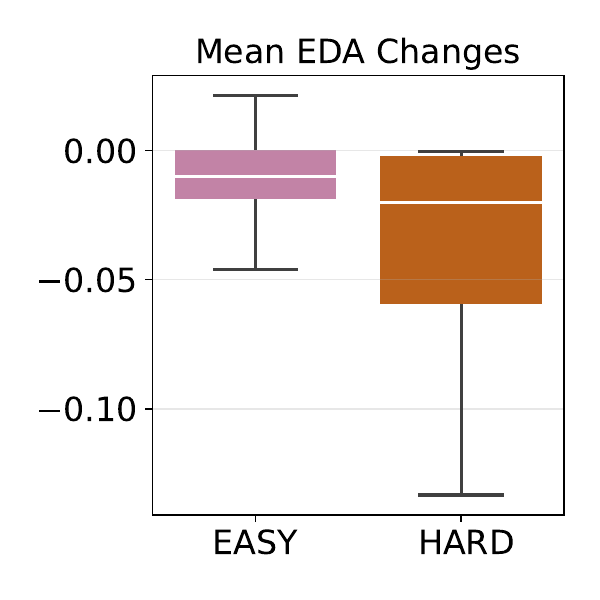}
    \end{subfigure}
    \begin{subfigure}[]{0.5\textwidth}
        \includegraphics[width=\textwidth]{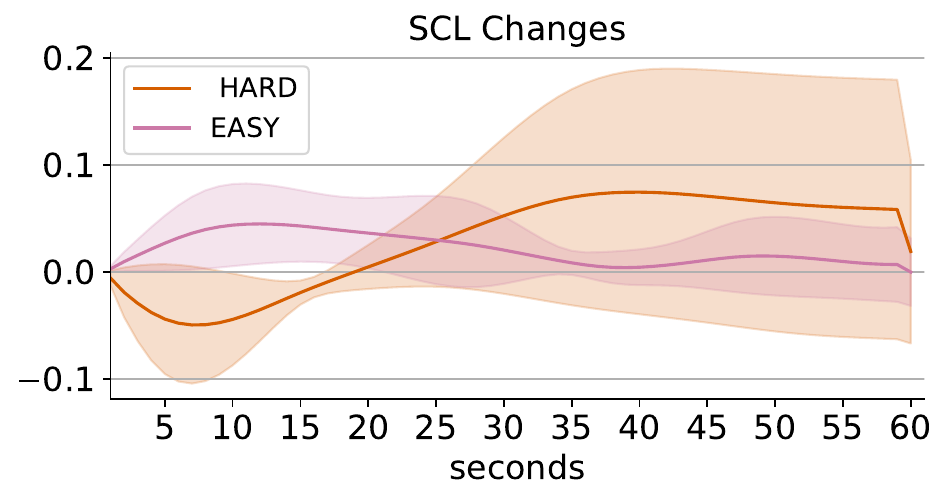}
    \end{subfigure}
    \caption{Preliminary EDA results ($N=7$) of the first minute on listening to Search Results on self-rated $easy$ and $hard$ topics. The left box plot represents the distribution of the average EDA changes, and the right line plot represents the changes of SCL over time.}
    \label{fig:preresult_eda}
\end{figure}

As illustrated in Figure~\ref{fig:preresult_eda}, when listening to the search results for the $easy$ topic, there is less individual difference in the changes of overall EDA value or the tonic value (i.e., SCL). However, the SCL exhibits much greater variability and fluctuation when participants are exposed to the $hard$ topic.
For the $hard$ topic, the SCL increases overall over time. This suggests the participants have increased arousal levels or feel more stress when absorbing the difficult information in the audio.
Conversely, the SCL overall decreases or remains mostly stable for the $easy$ topic.

\subsubsection{Summary}
These preliminary results indicate that observing comparable differences in users' consumption of information via audio (as is the case for SCS) is viable and worth exploring in more detail, since multi-modal signals (e.g., neural activity captured by EEG~\cite{bago2018fast, randall2014putting}) can provide insights into the activation of fast and slow thinking systems~\cite{kahneman2003perspective, kahneman2017thinking}.

By effectively combining the behavioral data (Section~\ref{subsec:tracking_behavioural}) and multiple signals from wearable devices (Section~\ref{subsec:multi-modal sensing}), we may eventually develop methods that can accurately identify user behavior and preferences in SCS scenarios. Our proposed methodology can potentially facilitate research in characterizing cognitive bias in SCS -- an area that requires further exploration.

\section{Challenges \& Ethical Considerations}
\label{subsec:challenges_ethics}

Physiological sensing has shown excellent performance in detecting emotion and specific cognitive activities. However, using sensors to detect complex constructs such as cognitive biases is not trivial. 
The need to use multiple wearable devices also raises various challenges and important ethical considerations that we discuss in this section.

\subsection{Planning}
Each method that captures behavioral and physiological responses provides data at different levels of granularity. Planning an SCS experiment involves carefully selecting appropriate measures.
We can use task responses to measure biases in the outcome of an information-processing activity, and transcripts and voice modulation can be used to measure biases in the process itself. However, the interactions in SCS are less structured compared to text-based search. As a result, analyzing behavioral data, such as SCS transcripts, requires an extensive qualitative approach (e.g., \cite{trippas2020towards}). This involves examining the data in detail and identifying patterns, themes, and insights, requiring more human effort. 
Physiological data reflects distinct changes that occur during SCS interaction stages. 
For example, reduced EDA is typically associated with engagement, while elevated EDA is connected to alertness. Peripherals can indicate emotional arousal and valence, but cognitive biases are complex. It is unclear whether cognitive biases have a directly observable impact on indicators that peripherals can measure.
Similarly, EEG data offers direct insight into neural activity with different analysis techniques, but each has distinct requirements for the experimental design, as discussed earlier. The lack of expertise relating to neuroscience can also be a challenge during study design. Since EEG data can include extensive noise, researchers must understand how to collect and interpret the data accurately.
Moreover, the duration of the activity is an often-overlooked factor that impacts on the reliability of physiological signal analysis. Typically, physiological sensing is used in longer activities (e.g., 40 minutes) as trends and patterns can unfold over extended periods. Conversely, SCS often involves short tasks of only a few minutes in length. Because of signal processing requirements, certain frequency features may become unavailable or distorted, especially those associated with high-frequency components in PPG data \cite{pham2021heart}.

We recommend that researchers develop a detailed data analysis plan during experiment design, and consider pre-registering the experiment.\footnote{{Preregistration is the process of specifying a research plan (including analysis) before carrying out a study, and submitting this to a registry. This allows for a clear distinction between exploratory research (such as looking for relationships in a data set) and confirmatory research (such as testing a predicted effect based on a clear hypothesis.}} They should have a sound understanding of the pros and cons of each data type, and corresponding analysis techniques.

\subsection{Interpretability} 
When a measured value has a distant relationship with the targeted construct, it is less valid, reliable, and objective \cite{riedl2014towards, wall2017warning}. Cognitive bias is a high-level, abstract construct. 
Researchers should break down the hypotheses relating to a cognitive bias into more concrete behaviors, such as engagement or cognitive load, and then further decompose these into direct indicators like skin conductance level or reaction time \cite{wall2017warning}. 

Physiological data typically provide indirect insights into the user experience. 
The information stemming from physiological data is embedded into a varying number of channels. 
Generally, the interpretation of data with fewer channels, such as peripherals and pupillary responses, are easier and more straightforward to analyze. In contrast, data with more channels, like EEG with 14 channels or more, requires complex analysis methods and multiple techniques.
Furthermore, examining changes in certain features extracted from data can help us better understand what has occurred, but there is a possibility of missing certain patterns. 
Multi-disciplinary expertise is needed to interpret the bio-mechanism behind such patterns \cite{schneegass2023future}. To this end, machine learning models with a large number of extracted features are commonly used \cite{akash2018trust, schmidt2019wearable, gwizdka2017temporal, d2010multimodal}. Deep learning models that decode entire signals can also reveal more intricate details \cite{GANAPATHY2020113571, wilson2021objective, Kangning2023Multimodal}, but risk making the analysis less interpretable.

\subsection{Sensitivity} 
Behavioral and physiological responses that can reveal more information are typically also more sensitive. 
In addition to controlled variables, other confounding variables, such as fatigue, interest, and health status could also affect responses.
The context, including specific activities such as speech, may also impact the data \cite{babaei2021critique, riedl2014towards}. Therefore, it is crucial to account for confounding variables both when designing the experiment and drawing conclusions from the results. Moreover, each signal requires unique pre-processing pipelines 
\cite{Bota2019}. The steps of pre-processing may overly trim (e.g., losing some patterns) or inadequately denoise (e.g., contaminating by artifacts) the data. 
When researching literature or conducting data-cleaning procedures, we recommend accounting for data sensitivity by keeping in mind the specific activities (e.g., speaking), contexts (e.g., mobile search~\cite{Kangning2023Multimodal}, collaborative information sharing~\cite{randall2014putting}), 
and devices (e.g., 14-channel~\cite{gwizdka2017temporal} or 1-channel~\cite{gwizdka2018inferring} EEG device) associated with the experiment. 

\subsection{Individual Difference} 
Everyone's body and brain can react and generate signals slightly differently due to genetics, age, health status, and past experiences. For instance, some people may have a calm disposition and are less prone to getting altered or excited. 
This variability means that a physiological response observed in one individual might not be identical or mean the same thing in another individual. Therefore, it is challenging to establish universal patterns, and interpreting physiological responses accurately often requires within-subject study designs, personalized or group-specific calibration and analysis.

\subsection{Data Collection} 

The data are collected across some barriers, e.g., skin (EDA) or skull (EGG). One key concern with this type of data is the potential for delay times and reliability issues. Specifically, there is a risk that the measured response to a given stimulus may be delayed and inadvertently counted as a response to the next stimulus. Besides, sensors used to collect data often require optimum contact with specific body areas (e.g., skin contact for EDA \cite{babaei2021critique}) to collect reliable data. It is important that the researchers are trained to familiarize themselves with the data and the devices before collecting the data.

\subsection{Ethical considerations}
Our proposed methodology introduces several potential risks that need to be considered. The sensitive nature of the data collected from wearable sensors (e.g., heart rate, or complex brain signals) requires a meticulous approach to data management, including collection, analysis, storage, and sharing. Data from devices like EEG can potentially compromise the privacy of participants \cite{yuste2023advocating} as they expose thoughts and emotions. Obtaining informed consent, and ensuring that participants are fully aware of the level of exposure, is essential.
Protocols such as the ones followed by \citet{arnau2021bed} for using EEG signals could be used. Recent studies have proposed EEG-based (i.e., biometric) login systems using different types of stimuli (emotional, cognitive, and/or physiological)~\cite{nakamura2017ear, chen2016high}. Researchers studying such complex signals need to be aware of the stimuli used for biometric purposes, and may need to refrain from collecting or storing similar stimuli in their experiments. 

In light of these factors, adhering to strict ethical standards, regulations, and protocols in handling such data is imperative. Furthermore, special attention needs to be paid when designing bias mitigation strategies using multi-modal signals for real-time content manipulation, as it may infringe an individual's cognitive liberty \cite{rainey2020brain}. More importantly, advances in this direction may require new regulations in order to minimize and control the adversarial use of content manipulation technology.
It is also crucial to account for individual variations, particularly among minority groups and those with brain injuries or neurological conditions (e.g., neurodiverse population). Experiments that consider these differences are necessary to achieve more accurate and representative results. Recent work by \citet{sitbon2023neurodiverseperspective} discusses practical approaches to include neurodiverse users in IR and related fields in detail.

\section{Authors' Positionality}

The paper discusses the perspectives which have been strongly influenced by the research, disciplinary background and personal views of our interdisciplinary author team. Our team includes computer science researchers in information retrieval, conversational search, human-computer interaction, and pervasive computing. 
Additionally, some of the authors have significantly influenced these perspectives from their work on exploring and conducting experiments around cognitive bias in either screen- or voice-based search, and their personal experience as members of the neurodiverse community.
By drawing from diverse perspectives, the authors acknowledge the complexities surrounding cognitive biases. The main goal of this paper is to support a comprehensive discussion on understanding and utilizing biases in SCS. We acknowledge the existing gap in including perspectives from minority groups, First Nations peoples~\cite{lewis2020indigenous,yunkaporta2023right}, or people with disabilities.
\section{Conclusions} 
\label{sec:conclusion}

This paper draws upon insights from disciplines including information-seeking, psychology, cognitive science, and wearable sensors. We highlight the under-explored area of cognitive biases in sophisticated voice-only systems like SCS and advocate for further investigation in this direction. We argue that instruments used to study cognitive biases in traditional web search are insufficient for this purpose, and propose research opportunities for further exploration in this area. Additionally, we envision and propose a general experimental approach for investigating cognitive biases in SCS, which researchers and practitioners can utilize. We report preliminary results from an exploratory experiment to demonstrate the feasibility and significance of utilizing neural activities and physiological signals for this purpose. We also acknowledge and identify the challenges of adopting our proposed approach, and point out some of the important ethical considerations that need to be made.


\begin{acks}
This research is partially supported by the \grantsponsor{ARC}{Australian Research Council}{https://www.arc.gov.au/} (\grantnum{ARC}{CE200100005}, \grantnum{ARC}{DE200100064}). 
We wish to extend our deep gratitude to Manasa Kalanadhabhatta, whose expertise in neuroscience was invaluable in ensuring the accuracy and clarity of the neuroscience-related sections.
\end{acks}

\newpage
\bibliographystyle{ACM-REFERENCE-FORMAT}
\bibliography{00_reference}


\begin{thebibliography}{138}


\ifx \showCODEN    \undefined \def \showCODEN     #1{\unskip}     \fi
\ifx \showDOI      \undefined \def \showDOI       #1{#1}\fi
\ifx \showISBNx    \undefined \def \showISBNx     #1{\unskip}     \fi
\ifx \showISBNxiii \undefined \def \showISBNxiii  #1{\unskip}     \fi
\ifx \showISSN     \undefined \def \showISSN      #1{\unskip}     \fi
\ifx \showLCCN     \undefined \def \showLCCN      #1{\unskip}     \fi
\ifx \shownote     \undefined \def \shownote      #1{#1}          \fi
\ifx \showarticletitle \undefined \def \showarticletitle #1{#1}   \fi
\ifx \showURL      \undefined \def \showURL       {\relax}        \fi
\providecommand\bibfield[2]{#2}
\providecommand\bibinfo[2]{#2}
\providecommand\natexlab[1]{#1}
\providecommand\showeprint[2][]{arXiv:#2}

\bibitem[Ackerman and Strickland(2022)]%
        {BRAIN2022}
\bibfield{author}{\bibinfo{person}{Evan Ackerman} {and} \bibinfo{person}{Eliza Strickland}.} \bibinfo{year}{2022}\natexlab{}.
\newblock \showarticletitle{Are You Ready for Workplace Brain Scanning?: Leveraging brain data will make workers happier and more productive, backers say}.
\newblock \bibinfo{journal}{\emph{IEEE Spectrum}} \bibinfo{volume}{59}, \bibinfo{number}{12} (\bibinfo{year}{2022}), \bibinfo{pages}{46--52}.
\newblock
\urldef\tempurl%
\url{https://doi.org/10.1109/MSPEC.2022.9976479}
\showDOI{\tempurl}


\bibitem[Ajjour et~al\mbox{.}(2019)]%
        {ajjour:2019a}
\bibfield{author}{\bibinfo{person}{Yamen Ajjour}, \bibinfo{person}{Henning Wachsmuth}, \bibinfo{person}{Johannes Kiesel}, \bibinfo{person}{Martin Potthast}, \bibinfo{person}{Matthias Hagen}, {and} \bibinfo{person}{Benno Stein}.} \bibinfo{year}{2019}\natexlab{}.
\newblock \showarticletitle{{Data Acquisition for Argument Search: The args.me corpus}}. In \bibinfo{booktitle}{\emph{42nd German Conference on Artificial Intelligence (KI 2019)}}, \bibfield{editor}{\bibinfo{person}{Christoph Benzm{\"u}ller} {and} \bibinfo{person}{Heiner Stuckenschmidt}} (Eds.). \bibinfo{publisher}{Springer}, \bibinfo{address}{Berlin Heidelberg New York}, \bibinfo{pages}{48--59}.
\newblock
\urldef\tempurl%
\url{https://doi.org/10.1007/978-3-030-30179-8\_4}
\showDOI{\tempurl}


\bibitem[Akash et~al\mbox{.}(2018)]%
        {akash2018trust}
\bibfield{author}{\bibinfo{person}{Kumar Akash}, \bibinfo{person}{Wan-Lin Hu}, \bibinfo{person}{Neera Jain}, {and} \bibinfo{person}{Tahira Reid}.} \bibinfo{year}{2018}\natexlab{}.
\newblock \showarticletitle{A Classification Model for Sensing Human Trust in Machines Using EEG and GSR}.
\newblock \bibinfo{journal}{\emph{ACM Trans. Interact. Intell. Syst.}} \bibinfo{volume}{8}, \bibinfo{number}{4}, Article \bibinfo{articleno}{27} (\bibinfo{date}{nov} \bibinfo{year}{2018}), \bibinfo{numpages}{20}~pages.
\newblock
\showISSN{2160-6455}
\urldef\tempurl%
\url{https://doi.org/10.1145/3132743}
\showDOI{\tempurl}


\bibitem[Alaofi et~al\mbox{.}(2022)]%
        {alaofi2022query}
\bibfield{author}{\bibinfo{person}{Marwah Alaofi}, \bibinfo{person}{Luke Gallagher}, \bibinfo{person}{Dana Mckay}, \bibinfo{person}{Lauren~L. Saling}, \bibinfo{person}{Mark Sanderson}, \bibinfo{person}{Falk Scholer}, \bibinfo{person}{Damiano Spina}, {and} \bibinfo{person}{Ryen~W. White}.} \bibinfo{year}{2022}\natexlab{}.
\newblock \showarticletitle{Where Do Queries Come From?}. In \bibinfo{booktitle}{\emph{Proceedings of the 45th International ACM SIGIR Conference on Research and Development in Information Retrieval}} (Madrid, Spain) \emph{(\bibinfo{series}{SIGIR '22})}. \bibinfo{publisher}{Association for Computing Machinery}, \bibinfo{address}{New York, NY, USA}, \bibinfo{pages}{2850–2862}.
\newblock
\showISBNx{9781450387323}
\urldef\tempurl%
\url{https://doi.org/10.1145/3477495.3531711}
\showDOI{\tempurl}


\bibitem[Aliannejadi and Trippas(2022)]%
        {alian2022conver}
\bibfield{author}{\bibinfo{person}{Mohammad Aliannejadi} {and} \bibinfo{person}{Johanne~R. Trippas}.} \bibinfo{year}{2022}\natexlab{}.
\newblock \showarticletitle{Conversational Information Seeking: Theory and Evaluation: CHIIR 2022 Half Day Tutorial}. In \bibinfo{booktitle}{\emph{Proceedings of the 2022 Conference on Human Information Interaction and Retrieval}} (Regensburg, Germany) \emph{(\bibinfo{series}{CHIIR '22})}. \bibinfo{publisher}{Association for Computing Machinery}, \bibinfo{address}{New York, NY, USA}, \bibinfo{pages}{365–366}.
\newblock
\showISBNx{9781450391863}
\urldef\tempurl%
\url{https://doi.org/10.1145/3498366.3505843}
\showDOI{\tempurl}


\bibitem[Allegretti et~al\mbox{.}(2015)]%
        {marco2015relevance}
\bibfield{author}{\bibinfo{person}{Marco Allegretti}, \bibinfo{person}{Yashar Moshfeghi}, \bibinfo{person}{Maria Hadjigeorgieva}, \bibinfo{person}{Frank~E. Pollick}, \bibinfo{person}{Joemon~M. Jose}, {and} \bibinfo{person}{Gabriella Pasi}.} \bibinfo{year}{2015}\natexlab{}.
\newblock \showarticletitle{When Relevance Judgement is Happening? An EEG-Based Study}. In \bibinfo{booktitle}{\emph{Proceedings of the 38th International ACM SIGIR Conference on Research and Development in Information Retrieval}} (Santiago, Chile) \emph{(\bibinfo{series}{SIGIR '15})}. \bibinfo{publisher}{Association for Computing Machinery}, \bibinfo{address}{New York, NY, USA}, \bibinfo{pages}{719–722}.
\newblock
\showISBNx{9781450336215}
\urldef\tempurl%
\url{https://doi.org/10.1145/2766462.2767811}
\showDOI{\tempurl}


\bibitem[Arnau-Gonz{\'a}lez et~al\mbox{.}(2021)]%
        {arnau2021bed}
\bibfield{author}{\bibinfo{person}{Pablo Arnau-Gonz{\'a}lez}, \bibinfo{person}{Stamos Katsigiannis}, \bibinfo{person}{Miguel Arevalillo-Herr{\'a}ez}, {and} \bibinfo{person}{Naeem Ramzan}.} \bibinfo{year}{2021}\natexlab{}.
\newblock \showarticletitle{BED: A New Data Set for EEG-based Biometrics}.
\newblock \bibinfo{journal}{\emph{IEEE Internet of Things Journal}} \bibinfo{volume}{8}, \bibinfo{number}{15} (\bibinfo{year}{2021}), \bibinfo{pages}{12219--12230}.
\newblock


\bibitem[Arya et~al\mbox{.}(2021)]%
        {ARYA2021100399}
\bibfield{author}{\bibinfo{person}{Resham Arya}, \bibinfo{person}{Jaiteg Singh}, {and} \bibinfo{person}{Ashok Kumar}.} \bibinfo{year}{2021}\natexlab{}.
\newblock \showarticletitle{A survey of multidisciplinary domains contributing to affective computing}.
\newblock \bibinfo{journal}{\emph{Computer Science Review}}  \bibinfo{volume}{40} (\bibinfo{year}{2021}), \bibinfo{pages}{100399}.
\newblock
\showISSN{1574-0137}
\urldef\tempurl%
\url{https://doi.org/10.1016/j.cosrev.2021.100399}
\showDOI{\tempurl}


\bibitem[Azemi et~al\mbox{.}(2023)]%
        {azemi2023biosignal}
\bibfield{author}{\bibinfo{person}{Erdrin Azemi}, \bibinfo{person}{Ali Moin}, \bibinfo{person}{Anuranjini Pragada}, \bibinfo{person}{Jean Hsiang-Chun Lu}, \bibinfo{person}{Victoria~M Powell}, \bibinfo{person}{Juri Minxha}, {and} \bibinfo{person}{Steven~P Hotelling}.} \bibinfo{year}{2023}\natexlab{}.
\newblock \bibinfo{title}{Biosignal sensing device using dynamic selection of electrodes}.
\newblock
\newblock
\newblock
\shownote{US Patent App. 18/094,841}.


\bibitem[Azzopardi(2021)]%
        {biasreview2021}
\bibfield{author}{\bibinfo{person}{Leif Azzopardi}.} \bibinfo{year}{2021}\natexlab{}.
\newblock \showarticletitle{Cognitive Biases in Search: A Review and Reflection of Cognitive Biases in Information Retrieval}. In \bibinfo{booktitle}{\emph{Proceedings of the 2021 Conference on Human Information Interaction and Retrieval}} (Canberra ACT, Australia) \emph{(\bibinfo{series}{CHIIR '21})}. \bibinfo{publisher}{Association for Computing Machinery}, \bibinfo{address}{New York, NY, USA}, \bibinfo{pages}{27–37}.
\newblock
\showISBNx{9781450380553}
\urldef\tempurl%
\url{https://doi.org/10.1145/3406522.3446023}
\showDOI{\tempurl}


\bibitem[Babaei et~al\mbox{.}(2021)]%
        {babaei2021critique}
\bibfield{author}{\bibinfo{person}{Ebrahim Babaei}, \bibinfo{person}{Benjamin Tag}, \bibinfo{person}{Tilman Dingler}, {and} \bibinfo{person}{Eduardo Velloso}.} \bibinfo{year}{2021}\natexlab{}.
\newblock \showarticletitle{A Critique of Electrodermal Activity Practices at CHI}. In \bibinfo{booktitle}{\emph{Proceedings of the 2021 CHI Conference on Human Factors in Computing Systems}} (Yokohama, Japan) \emph{(\bibinfo{series}{CHI '21})}. \bibinfo{publisher}{Association for Computing Machinery}, \bibinfo{address}{New York, NY, USA}, Article \bibinfo{articleno}{177}, \bibinfo{numpages}{14}~pages.
\newblock
\showISBNx{9781450380966}
\urldef\tempurl%
\url{https://doi.org/10.1145/3411764.3445370}
\showDOI{\tempurl}


\bibitem[Bago et~al\mbox{.}(2018)]%
        {bago2018fast}
\bibfield{author}{\bibinfo{person}{Bence Bago}, \bibinfo{person}{Darren Frey}, \bibinfo{person}{Julie Vidal}, \bibinfo{person}{Olivier Houdé}, \bibinfo{person}{Gregoire Borst}, {and} \bibinfo{person}{Wim {De Neys}}.} \bibinfo{year}{2018}\natexlab{}.
\newblock \showarticletitle{Fast and slow thinking: Electrophysiological evidence for early conflict sensitivity}.
\newblock \bibinfo{journal}{\emph{Neuropsychologia}}  \bibinfo{volume}{117} (\bibinfo{year}{2018}), \bibinfo{pages}{483--490}.
\newblock
\showISSN{0028-3932}
\urldef\tempurl%
\url{https://doi.org/10.1016/j.neuropsychologia.2018.07.017}
\showDOI{\tempurl}


\bibitem[B\'{e}n\'{e}dict et~al\mbox{.}(2023)]%
        {benedict2023generative}
\bibfield{author}{\bibinfo{person}{Garbiel B\'{e}n\'{e}dict}, \bibinfo{person}{Ruqing Zhang}, {and} \bibinfo{person}{Donald Metzler}.} \bibinfo{year}{2023}\natexlab{}.
\newblock \showarticletitle{Gen-IR@SIGIR 2023: The First Workshop on Generative Information Retrieval}. In \bibinfo{booktitle}{\emph{Proceedings of the 46th International ACM SIGIR Conference on Research and Development in Information Retrieval}} (Taipei, Taiwan) \emph{(\bibinfo{series}{SIGIR '23})}. \bibinfo{publisher}{Association for Computing Machinery}, \bibinfo{address}{New York, NY, USA}, \bibinfo{pages}{3460–3463}.
\newblock
\showISBNx{9781450394086}
\urldef\tempurl%
\url{https://doi.org/10.1145/3539618.3591923}
\showDOI{\tempurl}


\bibitem[Bigdeli et~al\mbox{.}(2023)]%
        {bigdeli2023understanding}
\bibfield{author}{\bibinfo{person}{Amin Bigdeli}, \bibinfo{person}{Negar Arabzadeh}, \bibinfo{person}{Shirin Seyedsalehi}, \bibinfo{person}{Morteza Zihayat}, {and} \bibinfo{person}{Ebrahim Bagheri}.} \bibinfo{year}{2023}\natexlab{}.
\newblock \showarticletitle{Understanding and Mitigating Gender Bias in Information Retrieval Systems}. In \bibinfo{booktitle}{\emph{European Conference on Information Retrieval}}. Springer, \bibinfo{pages}{315--323}.
\newblock


\bibitem[Bondarenko et~al\mbox{.}(2023)]%
        {bondarenko2023overview}
\bibfield{author}{\bibinfo{person}{Alexander Bondarenko}, \bibinfo{person}{Maik Fr{\"o}be}, \bibinfo{person}{Johannes Kiesel}, \bibinfo{person}{Ferdinand Schlatt}, \bibinfo{person}{Valentin Barriere}, \bibinfo{person}{Brian Ravenet}, \bibinfo{person}{L{\'e}o Hemamou}, \bibinfo{person}{Simon Luck}, \bibinfo{person}{Jan~Heinrich Reimer}, \bibinfo{person}{Benno Stein}, \bibinfo{person}{Martin Potthast}, {and} \bibinfo{person}{Matthias Hagen}.} \bibinfo{year}{2023}\natexlab{}.
\newblock \showarticletitle{Overview of Touch{\'e} 2023: Argument and Causal Retrieval}. In \bibinfo{booktitle}{\emph{Experimental IR Meets Multilinguality, Multimodality, and Interaction}}, \bibfield{editor}{\bibinfo{person}{Avi Arampatzis}, \bibinfo{person}{Evangelos Kanoulas}, \bibinfo{person}{Theodora Tsikrika}, \bibinfo{person}{Stefanos Vrochidis}, \bibinfo{person}{Anastasia Giachanou}, \bibinfo{person}{Dan Li}, \bibinfo{person}{Mohammad Aliannejadi}, \bibinfo{person}{Michalis Vlachos}, \bibinfo{person}{Guglielmo Faggioli}, {and} \bibinfo{person}{Nicola Ferro}} (Eds.). \bibinfo{publisher}{Springer Nature Switzerland}, \bibinfo{address}{Cham}, \bibinfo{pages}{507--530}.
\newblock
\showISBNx{978-3-031-42448-9}


\bibitem[Bondarenko et~al\mbox{.}(2022)]%
        {bondarenko2022overview}
\bibfield{author}{\bibinfo{person}{Alexander Bondarenko}, \bibinfo{person}{Maik Fr{\"o}be}, \bibinfo{person}{Johannes Kiesel}, \bibinfo{person}{Shahbaz Syed}, \bibinfo{person}{Timon Gurcke}, \bibinfo{person}{Meriem Beloucif}, \bibinfo{person}{Alexander Panchenko}, \bibinfo{person}{Chris Biemann}, \bibinfo{person}{Benno Stein}, \bibinfo{person}{Henning Wachsmuth}, {et~al\mbox{.}}} \bibinfo{year}{2022}\natexlab{}.
\newblock \showarticletitle{Overview of touch{\'e} 2022: argument retrieval}. In \bibinfo{booktitle}{\emph{International Conference of the Cross-Language Evaluation Forum for European Languages}}. Springer, \bibinfo{pages}{311--336}.
\newblock


\bibitem[Boonprakong et~al\mbox{.}(2023a)]%
        {boon2023bias}
\bibfield{author}{\bibinfo{person}{Nattapat Boonprakong}, \bibinfo{person}{Xiuge Chen}, \bibinfo{person}{Catherine Davey}, \bibinfo{person}{Benjamin Tag}, {and} \bibinfo{person}{Tilman Dingler}.} \bibinfo{year}{2023}\natexlab{a}.
\newblock \showarticletitle{Bias-Aware Systems: Exploring Indicators for the Occurrences of Cognitive Biases When Facing Different Opinions}. In \bibinfo{booktitle}{\emph{Proceedings of the 2023 CHI Conference on Human Factors in Computing Systems}} (Hamburg, Germany) \emph{(\bibinfo{series}{CHI '23})}. \bibinfo{publisher}{Association for Computing Machinery}, \bibinfo{address}{New York, NY, USA}, Article \bibinfo{articleno}{27}, \bibinfo{numpages}{19}~pages.
\newblock
\showISBNx{9781450394215}
\urldef\tempurl%
\url{https://doi.org/10.1145/3544548.3580917}
\showDOI{\tempurl}


\bibitem[Boonprakong et~al\mbox{.}(2023b)]%
        {Boonprakong2023exploringindicators}
\bibfield{author}{\bibinfo{person}{Nattapat Boonprakong}, \bibinfo{person}{Xiuge Chen}, \bibinfo{person}{Catherine Davey}, \bibinfo{person}{Benjamin Tag}, {and} \bibinfo{person}{Tilman Dingler}.} \bibinfo{year}{2023}\natexlab{b}.
\newblock \showarticletitle{Bias-Aware Systems: Exploring Indicators for the Occurrences of Cognitive Biases When Facing Different Opinions}. In \bibinfo{booktitle}{\emph{Proceedings of the 2023 CHI Conference on Human Factors in Computing Systems}} (Hamburg, Germany) \emph{(\bibinfo{series}{CHI '23})}. \bibinfo{publisher}{Association for Computing Machinery}, \bibinfo{address}{New York, NY, USA}, Article \bibinfo{articleno}{27}, \bibinfo{numpages}{19}~pages.
\newblock
\showISBNx{9781450394215}
\urldef\tempurl%
\url{https://doi.org/10.1145/3544548.3580917}
\showDOI{\tempurl}


\bibitem[Bos(2006)]%
        {bos2006eeg}
\bibfield{author}{\bibinfo{person}{Danny~Oude Bos}.} \bibinfo{year}{2006}\natexlab{}.
\newblock \showarticletitle{EEG-based emotion recognition}.
\newblock \bibinfo{journal}{\emph{The influence of visual and auditory stimuli}} \bibinfo{volume}{56}, \bibinfo{number}{3} (\bibinfo{year}{2006}), \bibinfo{pages}{1--17}.
\newblock


\bibitem[Bota et~al\mbox{.}(2019)]%
        {Bota2019}
\bibfield{author}{\bibinfo{person}{Patricia~J. Bota}, \bibinfo{person}{Chen Wang}, \bibinfo{person}{Ana~L.N. Fred}, {and} \bibinfo{person}{Hugo Placido~Da Silva}.} \bibinfo{year}{2019}\natexlab{}.
\newblock \showarticletitle{A Review, Current Challenges, and Future Possibilities on Emotion Recognition Using Machine Learning and Physiological Signals}.
\newblock \bibinfo{journal}{\emph{IEEE Access}}  \bibinfo{volume}{7} (\bibinfo{year}{2019}), \bibinfo{pages}{140990--141020}.
\newblock
\showISSN{21693536}
\urldef\tempurl%
\url{https://doi.org/10.1109/ACCESS.2019.2944001}
\showDOI{\tempurl}


\bibitem[Bragg et~al\mbox{.}(2018)]%
        {brag2018listen}
\bibfield{author}{\bibinfo{person}{Danielle Bragg}, \bibinfo{person}{Cynthia Bennett}, \bibinfo{person}{Katharina Reinecke}, {and} \bibinfo{person}{Richard Ladner}.} \bibinfo{year}{2018}\natexlab{}.
\newblock \showarticletitle{A Large Inclusive Study of Human Listening Rates}. In \bibinfo{booktitle}{\emph{Proceedings of the 2018 CHI Conference on Human Factors in Computing Systems}} (Montreal QC, Canada) \emph{(\bibinfo{series}{CHI '18})}. \bibinfo{publisher}{Association for Computing Machinery}, \bibinfo{address}{New York, NY, USA}, \bibinfo{pages}{1–12}.
\newblock
\showISBNx{9781450356206}
\urldef\tempurl%
\url{https://doi.org/10.1145/3173574.3174018}
\showDOI{\tempurl}


\bibitem[Buscher et~al\mbox{.}(2012)]%
        {buscher2012attentive}
\bibfield{author}{\bibinfo{person}{Georg Buscher}, \bibinfo{person}{Andreas Dengel}, \bibinfo{person}{Ralf Biedert}, {and} \bibinfo{person}{Ludger~V Elst}.} \bibinfo{year}{2012}\natexlab{}.
\newblock \showarticletitle{Attentive documents: Eye tracking as implicit feedback for information retrieval and beyond}.
\newblock \bibinfo{journal}{\emph{ACM Transactions on Interactive Intelligent Systems (TiiS)}} \bibinfo{volume}{1}, \bibinfo{number}{2} (\bibinfo{year}{2012}), \bibinfo{pages}{1--30}.
\newblock


\bibitem[Case et~al\mbox{.}(2005)]%
        {case2005avoiding}
\bibfield{author}{\bibinfo{person}{Donald~O Case}, \bibinfo{person}{James~E Andrews}, \bibinfo{person}{J~David Johnson}, {and} \bibinfo{person}{Suzanne~L Allard}.} \bibinfo{year}{2005}\natexlab{}.
\newblock \showarticletitle{Avoiding versus seeking: the relationship of information seeking to avoidance, blunting, coping, dissonance, and related concepts}.
\newblock \bibinfo{journal}{\emph{Journal of the Medical Library Association}} \bibinfo{volume}{93}, \bibinfo{number}{3} (\bibinfo{year}{2005}), \bibinfo{pages}{353}.
\newblock


\bibitem[Chadha et~al\mbox{.}(2012)]%
        {chadha2012listening}
\bibfield{author}{\bibinfo{person}{Monica Chadha}, \bibinfo{person}{Alex Avila}, {and} \bibinfo{person}{Homero Gil~de Z{\'u}{\~n}iga}.} \bibinfo{year}{2012}\natexlab{}.
\newblock \showarticletitle{Listening in: Building a profile of podcast users and analyzing their political participation}.
\newblock \bibinfo{journal}{\emph{Journal of Information Technology \& Politics}} \bibinfo{volume}{9}, \bibinfo{number}{4} (\bibinfo{year}{2012}), \bibinfo{pages}{388--401}.
\newblock


\bibitem[Chen et~al\mbox{.}(2016)]%
        {chen2016high}
\bibfield{author}{\bibinfo{person}{Yiyu Chen}, \bibinfo{person}{Ayalneh~Dessalegn Atnafu}, \bibinfo{person}{Isabella Schlattner}, \bibinfo{person}{Wendimagegn~Tariku Weldtsadik}, \bibinfo{person}{Myung-Cheol Roh}, \bibinfo{person}{Hyoung~Joong Kim}, \bibinfo{person}{Seong-Whan Lee}, \bibinfo{person}{Benjamin Blankertz}, {and} \bibinfo{person}{Siamac Fazli}.} \bibinfo{year}{2016}\natexlab{}.
\newblock \showarticletitle{A high-security EEG-based login system with RSVP stimuli and dry electrodes}.
\newblock \bibinfo{journal}{\emph{IEEE Transactions on Information Forensics and Security}} \bibinfo{volume}{11}, \bibinfo{number}{12} (\bibinfo{year}{2016}), \bibinfo{pages}{2635--2647}.
\newblock


\bibitem[Chuklin et~al\mbox{.}(2019)]%
        {chuklin2019using}
\bibfield{author}{\bibinfo{person}{Aleksandr Chuklin}, \bibinfo{person}{Aliaksei Severyn}, \bibinfo{person}{Johanne~R. Trippas}, \bibinfo{person}{Enrique Alfonseca}, \bibinfo{person}{Hanna Silen}, {and} \bibinfo{person}{Damiano Spina}.} \bibinfo{year}{2019}\natexlab{}.
\newblock \showarticletitle{Using Audio Transformations to Improve Comprehension in Voice Question Answering}. In \bibinfo{booktitle}{\emph{CLEF'19 Proceedigns of the Conference and Labs of the Evaluation Forum}}, \bibfield{editor}{\bibinfo{person}{Fabio Crestani}, \bibinfo{person}{Martin Braschler}, \bibinfo{person}{Jacques Savoy}, \bibinfo{person}{Andreas Rauber}, \bibinfo{person}{Henning Müller}, \bibinfo{person}{David~E. Losada}, \bibinfo{person}{Gundula Heinatz~Bürki}, \bibinfo{person}{Linda Cappellato}, {and} \bibinfo{person}{Nicola Ferro}} (Eds.). \bibinfo{publisher}{Springer International Publishing}, \bibinfo{address}{Cham}, \bibinfo{pages}{164--170}.
\newblock
\showISBNx{978-3-030-28577-7}


\bibitem[Clay et~al\mbox{.}(2013)]%
        {selective2013Review}
\bibfield{author}{\bibinfo{person}{Russ Clay}, \bibinfo{person}{Jessica~M. Barber}, {and} \bibinfo{person}{Natalie~J. Shook}.} \bibinfo{year}{2013}\natexlab{}.
\newblock \showarticletitle{Techniques for Measuring Selective Exposure: A Critical Review}.
\newblock \bibinfo{journal}{\emph{Communication Methods and Measures}} \bibinfo{volume}{7}, \bibinfo{number}{3-4} (\bibinfo{year}{2013}), \bibinfo{pages}{147--171}.
\newblock
\urldef\tempurl%
\url{https://doi.org/10.1080/19312458.2013.813925}
\showDOI{\tempurl}


\bibitem[Cole et~al\mbox{.}(2014)]%
        {cole2014task}
\bibfield{author}{\bibinfo{person}{Michael~J. Cole}, \bibinfo{person}{Chathra Hendahewa}, \bibinfo{person}{Nicholas~J. Belkin}, {and} \bibinfo{person}{Chirag Shah}.} \bibinfo{year}{2014}\natexlab{}.
\newblock \showarticletitle{Discrimination between Tasks with User Activity Patterns during Information Search}. In \bibinfo{booktitle}{\emph{Proceedings of the 37th International ACM SIGIR Conference on Research and Development in Information Retrieval}} (Gold Coast, Queensland, Australia) \emph{(\bibinfo{series}{SIGIR '14})}. \bibinfo{publisher}{Association for Computing Machinery}, \bibinfo{address}{New York, NY, USA}, \bibinfo{pages}{567–576}.
\newblock
\showISBNx{9781450322577}
\urldef\tempurl%
\url{https://doi.org/10.1145/2600428.2609591}
\showDOI{\tempurl}


\bibitem[Crestani and Du(2006)]%
        {crestani2006written}
\bibfield{author}{\bibinfo{person}{Fabio Crestani} {and} \bibinfo{person}{Heather Du}.} \bibinfo{year}{2006}\natexlab{}.
\newblock \showarticletitle{Written versus spoken queries: A qualitative and quantitative comparative analysis}.
\newblock \bibinfo{journal}{\emph{Journal of the American Society for Information Science and Technology}} \bibinfo{volume}{57}, \bibinfo{number}{7} (\bibinfo{year}{2006}), \bibinfo{pages}{881--890}.
\newblock


\bibitem[Daniel(2017)]%
        {kahneman2017thinking}
\bibfield{author}{\bibinfo{person}{Kahneman Daniel}.} \bibinfo{year}{2017}\natexlab{}.
\newblock \bibinfo{booktitle}{\emph{Thinking, Fast and Slow}}.
\newblock \bibinfo{publisher}{Farrar, Straus and Giroux}.
\newblock


\bibitem[Dingler et~al\mbox{.}(2022)]%
        {dingler2022method}
\bibfield{author}{\bibinfo{person}{Tilman Dingler}, \bibinfo{person}{Benjamin Tag}, \bibinfo{person}{David~A Eccles}, \bibinfo{person}{Niels Van~Berkel}, {and} \bibinfo{person}{Vassilis Kostakos}.} \bibinfo{year}{2022}\natexlab{}.
\newblock \showarticletitle{Method for Appropriating the Brief Implicit Association Test to Elicit Biases in Users}. In \bibinfo{booktitle}{\emph{Proceedings of the 2022 CHI Conference on Human Factors in Computing Systems}}. \bibinfo{pages}{1--16}.
\newblock


\bibitem[Draws et~al\mbox{.}(2021)]%
        {draws2021not}
\bibfield{author}{\bibinfo{person}{Tim Draws}, \bibinfo{person}{Nava Tintarev}, \bibinfo{person}{Ujwal Gadiraju}, \bibinfo{person}{Alessandro Bozzon}, {and} \bibinfo{person}{Benjamin Timmermans}.} \bibinfo{year}{2021}\natexlab{}.
\newblock \showarticletitle{This Is Not What We Ordered: Exploring Why Biased Search Result Rankings Affect User Attitudes on Debated Topics}. In \bibinfo{booktitle}{\emph{Proceedings of the 44th International ACM SIGIR Conference on Research and Development in Information Retrieval}}. \bibinfo{pages}{295--305}.
\newblock


\bibitem[Dubiel et~al\mbox{.}(2018)]%
        {dubiel2018investigating}
\bibfield{author}{\bibinfo{person}{Mateusz Dubiel}, \bibinfo{person}{Martin Halvey}, \bibinfo{person}{Leif Azzopardi}, {and} \bibinfo{person}{Sylvain Daronnat}.} \bibinfo{year}{2018}\natexlab{}.
\newblock \showarticletitle{Investigating how conversational search agents affect user's behaviour, performance and search experience}. In \bibinfo{booktitle}{\emph{The second international workshop on conversational approaches to information retrieval}}.
\newblock


\bibitem[D’mello and Graesser(2010)]%
        {d2010multimodal}
\bibfield{author}{\bibinfo{person}{Sidney~K D’mello} {and} \bibinfo{person}{Arthur Graesser}.} \bibinfo{year}{2010}\natexlab{}.
\newblock \showarticletitle{Multimodal semi-automated affect detection from conversational cues, gross body language, and facial features}.
\newblock \bibinfo{journal}{\emph{User Modeling and User-Adapted Interaction}}  \bibinfo{volume}{20} (\bibinfo{year}{2010}), \bibinfo{pages}{147--187}.
\newblock


\bibitem[Elkin and Leippe(1986)]%
        {elkin1986physiological}
\bibfield{author}{\bibinfo{person}{Roger~A Elkin} {and} \bibinfo{person}{Michael~R Leippe}.} \bibinfo{year}{1986}\natexlab{}.
\newblock \showarticletitle{Physiological arousal, dissonance, and attitude change: evidence for a dissonance-arousal link and a" don't remind me" effect.}
\newblock \bibinfo{journal}{\emph{Journal of personality and social psychology}} \bibinfo{volume}{51}, \bibinfo{number}{1} (\bibinfo{year}{1986}), \bibinfo{pages}{55}.
\newblock


\bibitem[Emotiv(2023)]%
        {Emotiv2023}
\bibfield{author}{\bibinfo{person}{Emotiv}.} \bibinfo{year}{2023}\natexlab{}.
\newblock \bibinfo{booktitle}{\emph{MN8 - 2 Channel EEG Earbuds}}.
\newblock
\urldef\tempurl%
\url{https://www.emotiv.com/mn8/}
\showURL{%
\tempurl}
\newblock
\shownote{Accessed: October 21, 2023}.


\bibitem[Empatica(2023)]%
        {Empatica2023}
\bibfield{author}{\bibinfo{person}{Empatica}.} \bibinfo{year}{2023}\natexlab{}.
\newblock \bibinfo{booktitle}{\emph{E4 wristband: Real-time physiological signals}}.
\newblock
\urldef\tempurl%
\url{https://www.empatica.com/research/e4/}
\showURL{%
\tempurl}
\newblock
\shownote{Accessed: October 21, 2023}.


\bibitem[Epstein and Robertson(2015)]%
        {epstein2015search}
\bibfield{author}{\bibinfo{person}{Robert Epstein} {and} \bibinfo{person}{Ronald~E Robertson}.} \bibinfo{year}{2015}\natexlab{}.
\newblock \showarticletitle{The search engine manipulation effect (SEME) and its possible impact on the outcomes of elections}.
\newblock \bibinfo{journal}{\emph{Proceedings of the National Academy of Sciences}} \bibinfo{volume}{112}, \bibinfo{number}{33} (\bibinfo{year}{2015}), \bibinfo{pages}{E4512--E4521}.
\newblock


\bibitem[Eugster et~al\mbox{.}(2016)]%
        {eugster2016natural}
\bibfield{author}{\bibinfo{person}{Manuel~JA Eugster}, \bibinfo{person}{Tuukka Ruotsalo}, \bibinfo{person}{Michiel~M Spap{\'e}}, \bibinfo{person}{Oswald Barral}, \bibinfo{person}{Niklas Ravaja}, \bibinfo{person}{Giulio Jacucci}, {and} \bibinfo{person}{Samuel Kaski}.} \bibinfo{year}{2016}\natexlab{}.
\newblock \showarticletitle{Natural brain-information interfaces: Recommending information by relevance inferred from human brain signals}.
\newblock \bibinfo{journal}{\emph{Scientific reports}} \bibinfo{volume}{6}, \bibinfo{number}{1} (\bibinfo{year}{2016}), \bibinfo{pages}{38580}.
\newblock


\bibitem[Federmeier et~al\mbox{.}(2020)]%
        {federmeier2020examining}
\bibfield{author}{\bibinfo{person}{Kara~D Federmeier}, \bibinfo{person}{Suzanne~R Jongman}, {and} \bibinfo{person}{Jakub~M Szewczyk}.} \bibinfo{year}{2020}\natexlab{}.
\newblock \showarticletitle{Examining the role of general cognitive skills in language processing: a window into complex cognition}.
\newblock \bibinfo{journal}{\emph{Current directions in psychological science}} \bibinfo{volume}{29}, \bibinfo{number}{6} (\bibinfo{year}{2020}), \bibinfo{pages}{575--582}.
\newblock


\bibitem[Francis and Love(2020)]%
        {francis2020listening}
\bibfield{author}{\bibinfo{person}{Alexander~L Francis} {and} \bibinfo{person}{Jordan Love}.} \bibinfo{year}{2020}\natexlab{}.
\newblock \showarticletitle{Listening effort: Are we measuring cognition or affect, or both?}
\newblock \bibinfo{journal}{\emph{Wiley Interdisciplinary Reviews: Cognitive Science}} \bibinfo{volume}{11}, \bibinfo{number}{1} (\bibinfo{year}{2020}), \bibinfo{pages}{e1514}.
\newblock


\bibitem[Ganapathy et~al\mbox{.}(2020)]%
        {GANAPATHY2020113571}
\bibfield{author}{\bibinfo{person}{Nagarajan Ganapathy}, \bibinfo{person}{Yedukondala~Rao Veeranki}, {and} \bibinfo{person}{Ramakrishnan Swaminathan}.} \bibinfo{year}{2020}\natexlab{}.
\newblock \showarticletitle{Convolutional neural network based emotion classification using electrodermal activity signals and time-frequency features}.
\newblock \bibinfo{journal}{\emph{Expert Systems with Applications}}  \bibinfo{volume}{159} (\bibinfo{year}{2020}), \bibinfo{pages}{113571}.
\newblock
\showISSN{0957-4174}
\urldef\tempurl%
\url{https://doi.org/10.1016/j.eswa.2020.113571}
\showDOI{\tempurl}


\bibitem[Gao and Shah(2020)]%
        {gao2020towards}
\bibfield{author}{\bibinfo{person}{Ruoyuan Gao} {and} \bibinfo{person}{Chirag Shah}.} \bibinfo{year}{2020}\natexlab{}.
\newblock \showarticletitle{Toward creating a fairer ranking in search engine results}.
\newblock \bibinfo{journal}{\emph{Information Processing \& Management}} \bibinfo{volume}{57}, \bibinfo{number}{1} (\bibinfo{year}{2020}), \bibinfo{pages}{102138}.
\newblock
\showISSN{0306-4573}
\urldef\tempurl%
\url{https://doi.org/10.1016/j.ipm.2019.102138}
\showDOI{\tempurl}


\bibitem[Gohsen et~al\mbox{.}(2023)]%
        {gohsen:2023a}
\bibfield{author}{\bibinfo{person}{Marcel Gohsen}, \bibinfo{person}{Johannes Kiesel}, \bibinfo{person}{Mariam Korashi}, \bibinfo{person}{Jan Ehlers}, {and} \bibinfo{person}{Benno Stein}.} \bibinfo{year}{2023}\natexlab{}.
\newblock \showarticletitle{{Guiding Oral Conversations: How to Nudge Users Towards Asking Questions?}}. In \bibinfo{booktitle}{\emph{ACM SIGIR Conference on Human Information Interaction and Retrieval (CHIIR 2023)}}. \bibinfo{publisher}{ACM}, \bibinfo{address}{New York}, \bibinfo{pages}{34--42}.
\newblock
\urldef\tempurl%
\url{https://doi.org/10.1145/3576840.3578291}
\showDOI{\tempurl}


\bibitem[Goodman and Mayhorn(2023)]%
        {GOODMAN2023103864}
\bibfield{author}{\bibinfo{person}{Kylie~L. Goodman} {and} \bibinfo{person}{Christopher~B. Mayhorn}.} \bibinfo{year}{2023}\natexlab{}.
\newblock \showarticletitle{It's not what you say but how you say it: Examining the influence of perceived voice assistant gender and pitch on trust and reliance}.
\newblock \bibinfo{journal}{\emph{Applied Ergonomics}}  \bibinfo{volume}{106} (\bibinfo{year}{2023}), \bibinfo{pages}{103864}.
\newblock
\showISSN{0003-6870}
\urldef\tempurl%
\url{https://doi.org/10.1016/j.apergo.2022.103864}
\showDOI{\tempurl}


\bibitem[Greenwald et~al\mbox{.}(1998)]%
        {greenwald1998measuring}
\bibfield{author}{\bibinfo{person}{Anthony~G Greenwald}, \bibinfo{person}{Debbie~E McGhee}, {and} \bibinfo{person}{Jordan~LK Schwartz}.} \bibinfo{year}{1998}\natexlab{}.
\newblock \showarticletitle{Measuring Individual Differences in Implicit Cognition: The Implicit Association Test}.
\newblock \bibinfo{journal}{\emph{Journal of personality and social psychology}} \bibinfo{volume}{74}, \bibinfo{number}{6} (\bibinfo{year}{1998}), \bibinfo{pages}{1464}.
\newblock


\bibitem[Guy(2018)]%
        {gud2018the}
\bibfield{author}{\bibinfo{person}{Ido Guy}.} \bibinfo{year}{2018}\natexlab{}.
\newblock \showarticletitle{The Characteristics of Voice Search: Comparing Spoken with Typed-in Mobile Web Search Queries}.
\newblock \bibinfo{journal}{\emph{ACM Trans. Inf. Syst.}} \bibinfo{volume}{36}, \bibinfo{number}{3}, Article \bibinfo{articleno}{30} (\bibinfo{date}{mar} \bibinfo{year}{2018}), \bibinfo{numpages}{28}~pages.
\newblock
\showISSN{1046-8188}
\urldef\tempurl%
\url{https://doi.org/10.1145/3182163}
\showDOI{\tempurl}


\bibitem[Gwizdka(2018)]%
        {gwizdka2018inferring}
\bibfield{author}{\bibinfo{person}{Jacek Gwizdka}.} \bibinfo{year}{2018}\natexlab{}.
\newblock \showarticletitle{Inferring Web Page Relevance Using Pupillometry and Single Channel EEG}. In \bibinfo{booktitle}{\emph{Information Systems and Neuroscience}}, \bibfield{editor}{\bibinfo{person}{Fred~D. Davis}, \bibinfo{person}{Ren{\'e} Riedl}, \bibinfo{person}{Jan vom Brocke}, \bibinfo{person}{Pierre-Majorique L{\'e}ger}, {and} \bibinfo{person}{Adriane~B. Randolph}} (Eds.), Vol.~\bibinfo{volume}{25}. \bibinfo{publisher}{Springer International Publishing}, \bibinfo{address}{Cham}, \bibinfo{pages}{175--183}.
\newblock
\showISBNx{978-3-319-67431-5}
\urldef\tempurl%
\url{https://doi.org/10.1007/978-3-319-67431-5_20}
\showDOI{\tempurl}


\bibitem[Gwizdka et~al\mbox{.}(2017)]%
        {gwizdka2017temporal}
\bibfield{author}{\bibinfo{person}{Jacek Gwizdka}, \bibinfo{person}{Rahilsadat Hosseini}, \bibinfo{person}{Michael Cole}, {and} \bibinfo{person}{Shouyi Wang}.} \bibinfo{year}{2017}\natexlab{}.
\newblock \showarticletitle{Temporal Dynamics of Eye-Tracking and EEG During Reading and Relevance Decisions}.
\newblock \bibinfo{journal}{\emph{Journal of the Association for Information Science and Technology}} \bibinfo{volume}{68}, \bibinfo{number}{10} (\bibinfo{year}{2017}), \bibinfo{pages}{2299--2312}.
\newblock


\bibitem[Harris(2019)]%
        {harris2019eyetracking}
\bibfield{author}{\bibinfo{person}{Christopher~G. Harris}.} \bibinfo{year}{2019}\natexlab{}.
\newblock \showarticletitle{Detecting Cognitive Bias in a Relevance Assessment Task Using an Eye Tracker}. In \bibinfo{booktitle}{\emph{Proceedings of the 11th ACM Symposium on Eye Tracking Research \& Applications}} (Denver, Colorado) \emph{(\bibinfo{series}{ETRA '19})}. \bibinfo{publisher}{Association for Computing Machinery}, \bibinfo{address}{New York, NY, USA}, Article \bibinfo{articleno}{36}, \bibinfo{numpages}{5}~pages.
\newblock
\showISBNx{9781450367097}
\urldef\tempurl%
\url{https://doi.org/10.1145/3314111.3319824}
\showDOI{\tempurl}


\bibitem[Hettiachchi et~al\mbox{.}(2020)]%
        {Hettiachchi2020}
\bibfield{author}{\bibinfo{person}{Danula Hettiachchi}, \bibinfo{person}{Zhanna Sarsenbayeva}, \bibinfo{person}{Fraser Allison}, \bibinfo{person}{Niels van Berkel}, \bibinfo{person}{Tilman Dingler}, \bibinfo{person}{Gabriele Marini}, \bibinfo{person}{Vassilis Kostakos}, {and} \bibinfo{person}{Jorge Goncalves}.} \bibinfo{year}{2020}\natexlab{}.
\newblock \showarticletitle{"Hi! I am the Crowd Tasker" Crowdsourcing through Digital Voice Assistants}. In \bibinfo{booktitle}{\emph{Proceedings of the 2020 CHI Conference on Human Factors in Computing Systems}} (Honolulu, HI, USA) \emph{(\bibinfo{series}{CHI '20})}. \bibinfo{publisher}{Association for Computing Machinery}, \bibinfo{address}{New York, NY, USA}, \bibinfo{pages}{1–14}.
\newblock
\showISBNx{9781450367080}
\urldef\tempurl%
\url{https://doi.org/10.1145/3313831.3376320}
\showDOI{\tempurl}


\bibitem[Ji et~al\mbox{.}(2024)]%
        {ji2024characterizing}
\bibfield{author}{\bibinfo{person}{Kaixin Ji}, \bibinfo{person}{Danula Hettiachchi}, \bibinfo{person}{Flora~D. Salim}, \bibinfo{person}{Falk Scholer}, {and} \bibinfo{person}{Damiano Spina}.} \bibinfo{year}{2024}\natexlab{}.
\newblock \showarticletitle{Characterizing Information Seeking Processes with Multiple Physiological Signals}. In \bibinfo{booktitle}{\emph{Proceedings of the 47th International ACM SIGIR Conference on Research and Development in Information Retrieval}} (Washington, DC, USA) \emph{(\bibinfo{series}{SIGIR '24})}. \bibinfo{publisher}{Association for Computing Machinery}, \bibinfo{address}{New York, NY, USA}, \bibinfo{numpages}{12}~pages.
\newblock
\urldef\tempurl%
\url{https://doi.org/10.1145/3626772.3657793}
\showDOI{\tempurl}


\bibitem[Ji et~al\mbox{.}(2023)]%
        {ji2023towards}
\bibfield{author}{\bibinfo{person}{Kaixin Ji}, \bibinfo{person}{Damiano Spina}, \bibinfo{person}{Danula Hettiachchi}, \bibinfo{person}{Flora~Dylis Salim}, {and} \bibinfo{person}{Falk Scholer}.} \bibinfo{year}{2023}\natexlab{}.
\newblock \showarticletitle{Towards Detecting Tonic Information Processing Activities with Physiological Data}. In \bibinfo{booktitle}{\emph{Adjunct Proceedings of the 2023 ACM International Joint Conference on Pervasive and Ubiquitous Computing \& the 2023 ACM International Symposium on Wearable Computing}} (Cancun, Quintana Roo, Mexico) \emph{(\bibinfo{series}{UbiComp/ISWC '23 Adjunct})}. \bibinfo{publisher}{Association for Computing Machinery}, \bibinfo{address}{New York, NY, USA}, \bibinfo{numpages}{5}~pages.
\newblock
\showISBNx{979-8-4007-0200-6/23/10}
\urldef\tempurl%
\url{https://doi.org/10.1145/3594739.3610679}
\showDOI{\tempurl}


\bibitem[Jiang et~al\mbox{.}(2020)]%
        {jiang2020believe}
\bibfield{author}{\bibinfo{person}{Xiaoming Jiang}, \bibinfo{person}{Kira Gossack-Keenan}, {and} \bibinfo{person}{Marc~D Pell}.} \bibinfo{year}{2020}\natexlab{}.
\newblock \showarticletitle{To believe or not to believe? How voice and accent information in speech alter listener impressions of trust}.
\newblock \bibinfo{journal}{\emph{Quarterly Journal of Experimental Psychology}} \bibinfo{volume}{73}, \bibinfo{number}{1} (\bibinfo{year}{2020}), \bibinfo{pages}{55--79}.
\newblock


\bibitem[Jimenez-Molina et~al\mbox{.}(2018)]%
        {jimenez2018using}
\bibfield{author}{\bibinfo{person}{Angel Jimenez-Molina}, \bibinfo{person}{Cristian Retamal}, {and} \bibinfo{person}{Hernan Lira}.} \bibinfo{year}{2018}\natexlab{}.
\newblock \showarticletitle{Using psychophysiological sensors to assess mental workload during web browsing}.
\newblock \bibinfo{journal}{\emph{Sensors}} \bibinfo{volume}{18}, \bibinfo{number}{2} (\bibinfo{year}{2018}), \bibinfo{pages}{458}.
\newblock


\bibitem[Kahneman(2003)]%
        {kahneman2003perspective}
\bibfield{author}{\bibinfo{person}{Daniel Kahneman}.} \bibinfo{year}{2003}\natexlab{}.
\newblock \showarticletitle{A Perspective on Judgment and Choice: Mapping Bounded Rationality.}
\newblock \bibinfo{journal}{\emph{American Psychologist}} \bibinfo{volume}{58}, \bibinfo{number}{9} (\bibinfo{year}{2003}), \bibinfo{pages}{697}.
\newblock
\urldef\tempurl%
\url{https://doi.org/10.1037/0003-066X.58.9.697}
\showDOI{\tempurl}


\bibitem[Kahneman et~al\mbox{.}(2021)]%
        {kahneman2021noise}
\bibfield{author}{\bibinfo{person}{Daniel Kahneman}, \bibinfo{person}{Olivier Sibony}, {and} \bibinfo{person}{Cass~R Sunstein}.} \bibinfo{year}{2021}\natexlab{}.
\newblock \bibinfo{booktitle}{\emph{Noises: A Flaw in Human Judgement}}.
\newblock \bibinfo{publisher}{Little, Brown Spark}.
\newblock
\showISBNx{978-0-00-830899-5}


\bibitem[Karat et~al\mbox{.}(1999)]%
        {karat1999patterns}
\bibfield{author}{\bibinfo{person}{Clare-Marie Karat}, \bibinfo{person}{Christine Halverson}, \bibinfo{person}{Daniel Horn}, {and} \bibinfo{person}{John Karat}.} \bibinfo{year}{1999}\natexlab{}.
\newblock \showarticletitle{Patterns of entry and correction in large vocabulary continuous speech recognition systems}. In \bibinfo{booktitle}{\emph{Proceedings of the SIGCHI Conference on Human Factors in Computing Systems}} (Pittsburgh, Pennsylvania, USA) \emph{(\bibinfo{series}{CHI '99})}. \bibinfo{publisher}{Association for Computing Machinery}, \bibinfo{address}{New York, NY, USA}, \bibinfo{pages}{568–575}.
\newblock
\showISBNx{0201485591}
\urldef\tempurl%
\url{https://doi.org/10.1145/302979.303160}
\showDOI{\tempurl}


\bibitem[Khofman(2023)]%
        {khofman2023exploring}
\bibfield{author}{\bibinfo{person}{Anna Khofman}.} \bibinfo{year}{2023}\natexlab{}.
\newblock \bibinfo{title}{Exploring cognitive biases in voice-based virtual assistants}.
\newblock
\newblock


\bibitem[Kiesel et~al\mbox{.}(2018)]%
        {kiesel2018voiceclarifying}
\bibfield{author}{\bibinfo{person}{Johannes Kiesel}, \bibinfo{person}{Arefeh Bahrami}, \bibinfo{person}{Benno Stein}, \bibinfo{person}{Avishek Anand}, {and} \bibinfo{person}{Matthias Hagen}.} \bibinfo{year}{2018}\natexlab{}.
\newblock \showarticletitle{Toward Voice Query Clarification}. In \bibinfo{booktitle}{\emph{The 41st International ACM SIGIR Conference on Research \& Development in Information Retrieval}} (Ann Arbor, MI, USA) \emph{(\bibinfo{series}{SIGIR '18})}. \bibinfo{publisher}{Association for Computing Machinery}, \bibinfo{address}{New York, NY, USA}, \bibinfo{pages}{1257–1260}.
\newblock
\showISBNx{9781450356572}
\urldef\tempurl%
\url{https://doi.org/10.1145/3209978.3210160}
\showDOI{\tempurl}


\bibitem[Kiesel et~al\mbox{.}(2019)]%
        {kiesel2019falsememories}
\bibfield{author}{\bibinfo{person}{Johannes Kiesel}, \bibinfo{person}{Arefeh Bahrami}, \bibinfo{person}{Benno Stein}, \bibinfo{person}{Avishek Anand}, {and} \bibinfo{person}{Matthias Hagen}.} \bibinfo{year}{2019}\natexlab{}.
\newblock \showarticletitle{Clarifying False Memories in Voice-Based Search}. In \bibinfo{booktitle}{\emph{Proceedings of the 2019 Conference on Human Information Interaction and Retrieval}} (Glasgow, Scotland UK) \emph{(\bibinfo{series}{CHIIR '19})}. \bibinfo{publisher}{Association for Computing Machinery}, \bibinfo{address}{New York, NY, USA}, \bibinfo{pages}{331–335}.
\newblock
\showISBNx{9781450360258}
\urldef\tempurl%
\url{https://doi.org/10.1145/3295750.3298961}
\showDOI{\tempurl}


\bibitem[Kiesel et~al\mbox{.}(2021)]%
        {Kiesel2021conv}
\bibfield{author}{\bibinfo{person}{Johannes Kiesel}, \bibinfo{person}{Damiano Spina}, \bibinfo{person}{Henning Wachsmuth}, {and} \bibinfo{person}{Benno Stein}.} \bibinfo{year}{2021}\natexlab{}.
\newblock \showarticletitle{The Meant, the Said, and the Understood: Conversational Argument Search and Cognitive Biases}. In \bibinfo{booktitle}{\emph{Proceedings of the 3rd Conference on Conversational User Interfaces}} (Bilbao (online), Spain) \emph{(\bibinfo{series}{CUI '21})}. \bibinfo{publisher}{Association for Computing Machinery}, \bibinfo{address}{New York, NY, USA}, Article \bibinfo{articleno}{20}, \bibinfo{numpages}{5}~pages.
\newblock
\showISBNx{9781450389983}
\urldef\tempurl%
\url{https://doi.org/10.1145/3469595.3469615}
\showDOI{\tempurl}


\bibitem[Kreibig(2010)]%
        {KREIBIG2010394}
\bibfield{author}{\bibinfo{person}{Sylvia~D. Kreibig}.} \bibinfo{year}{2010}\natexlab{}.
\newblock \showarticletitle{Autonomic nervous system activity in emotion: A review}.
\newblock \bibinfo{journal}{\emph{Biological Psychology}} \bibinfo{volume}{84}, \bibinfo{number}{3} (\bibinfo{year}{2010}), \bibinfo{pages}{394--421}.
\newblock
\showISSN{0301-0511}
\urldef\tempurl%
\url{https://doi.org/10.1016/j.biopsycho.2010.03.010}
\showDOI{\tempurl}
\newblock
\shownote{The biopsychology of emotion: Current theoretical and empirical perspectives}.


\bibitem[Kumar and Bhuvaneswari(2012)]%
        {KUMAR20122525}
\bibfield{author}{\bibinfo{person}{J.~Satheesh Kumar} {and} \bibinfo{person}{P. Bhuvaneswari}.} \bibinfo{year}{2012}\natexlab{}.
\newblock \showarticletitle{Analysis of Electroencephalography (EEG) Signals and Its Categorization–A Study}.
\newblock \bibinfo{journal}{\emph{Procedia Engineering}}  \bibinfo{volume}{38} (\bibinfo{year}{2012}), \bibinfo{pages}{2525--2536}.
\newblock
\showISSN{1877-7058}
\urldef\tempurl%
\url{https://doi.org/10.1016/j.proeng.2012.06.298}
\showDOI{\tempurl}
\newblock
\shownote{International Conference on Modelling Optimization and Computing}.


\bibitem[Labs(2023)]%
        {PupilLabs2023}
\bibfield{author}{\bibinfo{person}{Pupil Labs}.} \bibinfo{year}{2023}\natexlab{}.
\newblock \bibinfo{booktitle}{\emph{Neon: Eye tracking for research and beyond}}.
\newblock
\urldef\tempurl%
\url{https://pupil-labs.com/products/}
\showURL{%
\tempurl}
\newblock
\shownote{Accessed: October 21, 2023}.


\bibitem[Lee et~al\mbox{.}(2020)]%
        {sunok2020usermentalmodel}
\bibfield{author}{\bibinfo{person}{Sunok Lee}, \bibinfo{person}{Minji Cho}, {and} \bibinfo{person}{Sangsu Lee}.} \bibinfo{year}{2020}\natexlab{}.
\newblock \showarticletitle{What If Conversational Agents Became Invisible? Comparing Users' Mental Models According to Physical Entity of AI Speaker}.
\newblock \bibinfo{journal}{\emph{Proc. ACM Interact. Mob. Wearable Ubiquitous Technol.}} \bibinfo{volume}{4}, \bibinfo{number}{3}, Article \bibinfo{articleno}{88} (\bibinfo{date}{sep} \bibinfo{year}{2020}), \bibinfo{numpages}{24}~pages.
\newblock
\urldef\tempurl%
\url{https://doi.org/10.1145/3411840}
\showDOI{\tempurl}


\bibitem[Leong{\'o}mez et~al\mbox{.}(2021)]%
        {leongomez2021voice}
\bibfield{author}{\bibinfo{person}{Juan~David Leong{\'o}mez}, \bibinfo{person}{Katarzyna Pisanski}, \bibinfo{person}{David Reby}, \bibinfo{person}{Disa Sauter}, \bibinfo{person}{Nadine Lavan}, \bibinfo{person}{Marcus Perlman}, {and} \bibinfo{person}{Jaroslava Varella~Valentova}.} \bibinfo{year}{2021}\natexlab{}.
\newblock \bibinfo{title}{Voice modulation: from origin and mechanism to social impact}.
\newblock , \bibinfo{numpages}{20200386}~pages.
\newblock


\bibitem[Lewandowsky et~al\mbox{.}(2012)]%
        {lewandowsky13schwarz}
\bibfield{author}{\bibinfo{person}{Stephan Lewandowsky}, \bibinfo{person}{KH Ecker~Ullrich}, {and} \bibinfo{person}{M Seifert~Colleen}.} \bibinfo{year}{2012}\natexlab{}.
\newblock \showarticletitle{Misinformation and Its Correction: Continued Influence and Successful Debiasing}.
\newblock \bibinfo{journal}{\emph{Psychological Science in the Public Interest}} \bibinfo{volume}{13}, \bibinfo{number}{3} (\bibinfo{year}{2012}), \bibinfo{pages}{106--131}.
\newblock
\urldef\tempurl%
\url{https://doi.org/10.1177/1529100612451018}
\showDOI{\tempurl}


\bibitem[Lewis et~al\mbox{.}(2020a)]%
        {lewis2020indigenous}
\bibfield{author}{\bibinfo{person}{Jason~Edward Lewis}, \bibinfo{person}{Angie Abdilla}, \bibinfo{person}{Noelani Arista}, \bibinfo{person}{Kaipulaumakaniolono Baker}, \bibinfo{person}{Scott Benesiinaabandan}, \bibinfo{person}{Michelle Brown}, \bibinfo{person}{Melanie Cheung}, \bibinfo{person}{Meredith Coleman}, \bibinfo{person}{Ashley Cordes}, \bibinfo{person}{Joel Davison}, \bibinfo{person}{K{\=u}pono Duncan}, \bibinfo{person}{Sergio Garzon}, \bibinfo{person}{D.~Fox Harrell}, \bibinfo{person}{Peter-Lucas Jones}, \bibinfo{person}{Kekuhi Kealiikanakaoleohaililani}, \bibinfo{person}{Megan Kelleher}, \bibinfo{person}{Suzanne Kite}, \bibinfo{person}{Olin Lagon}, \bibinfo{person}{Jason Leigh}, \bibinfo{person}{Maroussia Levesque}, \bibinfo{person}{Keoni Mahelona}, \bibinfo{person}{Caleb Moses}, \bibinfo{person}{Isaac~('Ika'aka) Nahuewai}, \bibinfo{person}{Kari Noe}, \bibinfo{person}{Danielle Olson}, \bibinfo{person}{'{\=O}iwi Parker~Jones}, \bibinfo{person}{Caroline Running~Wolf}, \bibinfo{person}{Michael
  Running~Wolf}, \bibinfo{person}{Marlee Silva}, \bibinfo{person}{Skawennati Fragnito}, {and} \bibinfo{person}{H{\=e}mi Whaanga}.} \bibinfo{year}{2020}\natexlab{a}.
\newblock \showarticletitle{Indigenous {{Protocol}} and {{Artificial Intelligence Position Paper}}}.
\newblock  (\bibinfo{year}{2020}).
\newblock
\urldef\tempurl%
\url{https://doi.org/10.11573/SPECTRUM.LIBRARY.CONCORDIA.CA.00986506}
\showDOI{\tempurl}


\bibitem[Lewis et~al\mbox{.}(2020b)]%
        {lewis2020retrieval}
\bibfield{author}{\bibinfo{person}{Patrick Lewis}, \bibinfo{person}{Ethan Perez}, \bibinfo{person}{Aleksandra Piktus}, \bibinfo{person}{Fabio Petroni}, \bibinfo{person}{Vladimir Karpukhin}, \bibinfo{person}{Naman Goyal}, \bibinfo{person}{Heinrich K{\"u}ttler}, \bibinfo{person}{Mike Lewis}, \bibinfo{person}{Wen-tau Yih}, \bibinfo{person}{Tim Rockt{\"a}schel}, {et~al\mbox{.}}} \bibinfo{year}{2020}\natexlab{b}.
\newblock \showarticletitle{Retrieval-augmented generation for knowledge-intensive nlp tasks}.
\newblock \bibinfo{journal}{\emph{Advances in Neural Information Processing Systems}}  \bibinfo{volume}{33} (\bibinfo{year}{2020}), \bibinfo{pages}{9459--9474}.
\newblock


\bibitem[Lewis et~al\mbox{.}(2020c)]%
        {lewis200rag}
\bibfield{author}{\bibinfo{person}{Patrick Lewis}, \bibinfo{person}{Ethan Perez}, \bibinfo{person}{Aleksandra Piktus}, \bibinfo{person}{Fabio Petroni}, \bibinfo{person}{Vladimir Karpukhin}, \bibinfo{person}{Naman Goyal}, \bibinfo{person}{Heinrich K\"{u}ttler}, \bibinfo{person}{Mike Lewis}, \bibinfo{person}{Wen-tau Yih}, \bibinfo{person}{Tim Rockt\"{a}schel}, \bibinfo{person}{Sebastian Riedel}, {and} \bibinfo{person}{Douwe Kiela}.} \bibinfo{year}{2020}\natexlab{c}.
\newblock \showarticletitle{Retrieval-Augmented Generation for Knowledge-Intensive NLP Tasks}. In \bibinfo{booktitle}{\emph{Proceedings of the 34th International Conference on Neural Information Processing Systems}} (Vancouver, BC, Canada) \emph{(\bibinfo{series}{NIPS'20})}. \bibinfo{publisher}{Curran Associates Inc.}, \bibinfo{address}{Red Hook, NY, USA}, Article \bibinfo{articleno}{793}, \bibinfo{numpages}{16}~pages.
\newblock
\showISBNx{9781713829546}


\bibitem[Liao et~al\mbox{.}(2023)]%
        {liao2023proactive}
\bibfield{author}{\bibinfo{person}{Lizi Liao}, \bibinfo{person}{Grace~Hui Yang}, {and} \bibinfo{person}{Chirag Shah}.} \bibinfo{year}{2023}\natexlab{}.
\newblock \showarticletitle{Proactive Conversational Agents in the Post-ChatGPT World}. In \bibinfo{booktitle}{\emph{Proceedings of the 46th International ACM SIGIR Conference on Research and Development in Information Retrieval}} (Taipei, Taiwan) \emph{(\bibinfo{series}{SIGIR '23})}. \bibinfo{publisher}{Association for Computing Machinery}, \bibinfo{address}{New York, NY, USA}, \bibinfo{pages}{3452–3455}.
\newblock
\showISBNx{9781450394086}
\urldef\tempurl%
\url{https://doi.org/10.1145/3539618.3594250}
\showDOI{\tempurl}


\bibitem[Lindgren(2023)]%
        {lindgren2023intimacy}
\bibfield{author}{\bibinfo{person}{Mia Lindgren}.} \bibinfo{year}{2023}\natexlab{}.
\newblock \showarticletitle{Intimacy and emotions in podcast journalism: A study of award-winning Australian and British podcasts}.
\newblock \bibinfo{journal}{\emph{Journalism Practice}} \bibinfo{volume}{17}, \bibinfo{number}{4} (\bibinfo{year}{2023}), \bibinfo{pages}{704--719}.
\newblock


\bibitem[Luck and Kappenman(2016)]%
        {luck_kappenman_2016}
\bibfield{author}{\bibinfo{person}{Steven~J. Luck} {and} \bibinfo{person}{Emily~S. Kappenman}.} \bibinfo{year}{2016}\natexlab{}.
\newblock \bibinfo{booktitle}{\emph{Electroencephalography and Event-Related Brain Potentials} (\bibinfo{edition}{4} ed.)}.
\newblock \bibinfo{publisher}{Cambridge University Press}, \bibinfo{pages}{74–100}.
\newblock
\urldef\tempurl%
\url{https://doi.org/10.1017/9781107415782.005}
\showDOI{\tempurl}


\bibitem[Marois et~al\mbox{.}(2019)]%
        {marois2019auditory}
\bibfield{author}{\bibinfo{person}{Alexandre Marois}, \bibinfo{person}{John~E Marsh}, {and} \bibinfo{person}{Fran{\c{c}}ois Vachon}.} \bibinfo{year}{2019}\natexlab{}.
\newblock \showarticletitle{Is auditory distraction by changing-state and deviant sounds underpinned by the same mechanism? Evidence from pupillometry}.
\newblock \bibinfo{journal}{\emph{Biological Psychology}}  \bibinfo{volume}{141} (\bibinfo{year}{2019}), \bibinfo{pages}{64--74}.
\newblock


\bibitem[Martínez-Maldonado et~al\mbox{.}(2020)]%
        {MARTINEZMALDONADO202046}
\bibfield{author}{\bibinfo{person}{Andrés Martínez-Maldonado}, \bibinfo{person}{Rosa Jurado-Barba}, \bibinfo{person}{Ana Sion}, \bibinfo{person}{Isabel Domínguez-Centeno}, \bibinfo{person}{Gabriela Castillo-Parra}, \bibinfo{person}{Julio Prieto-Montalvo}, {and} \bibinfo{person}{Gabriel Rubio}.} \bibinfo{year}{2020}\natexlab{}.
\newblock \showarticletitle{Brain functional connectivity after cognitive-bias modification and behavioral changes in abstinent alcohol-use disorder patients}.
\newblock \bibinfo{journal}{\emph{International Journal of Psychophysiology}}  \bibinfo{volume}{154} (\bibinfo{year}{2020}), \bibinfo{pages}{46--58}.
\newblock
\showISSN{0167-8760}
\urldef\tempurl%
\url{https://doi.org/10.1016/j.ijpsycho.2019.10.004}
\showDOI{\tempurl}
\newblock
\shownote{Neurophysiologic impact of non-pharmacologic interventions for cognition}.


\bibitem[Maxwell and Azzopardi(2016)]%
        {maxwell2016agents}
\bibfield{author}{\bibinfo{person}{David Maxwell} {and} \bibinfo{person}{Leif Azzopardi}.} \bibinfo{year}{2016}\natexlab{}.
\newblock \showarticletitle{Agents, simulated users and humans: An analysis of performance and behaviour}. In \bibinfo{booktitle}{\emph{Proceedings of the 25th ACM international on conference on information and knowledge management}}. \bibinfo{pages}{731--740}.
\newblock


\bibitem[Melumad(2023)]%
        {melumad2023vocalizing}
\bibfield{author}{\bibinfo{person}{Shiri Melumad}.} \bibinfo{year}{2023}\natexlab{}.
\newblock \showarticletitle{Vocalizing search: How voice technologies alter consumer search processes and satisfaction}.
\newblock \bibinfo{journal}{\emph{Journal of Consumer Research}} (\bibinfo{year}{2023}), \bibinfo{pages}{ucad009}.
\newblock


\bibitem[Metzger and Flanagin(2013)]%
        {metzger2013credibility}
\bibfield{author}{\bibinfo{person}{Miriam~J Metzger} {and} \bibinfo{person}{Andrew~J Flanagin}.} \bibinfo{year}{2013}\natexlab{}.
\newblock \showarticletitle{Credibility and trust of information in online environments: The use of cognitive heuristics}.
\newblock \bibinfo{journal}{\emph{Journal of pragmatics}}  \bibinfo{volume}{59} (\bibinfo{year}{2013}), \bibinfo{pages}{210--220}.
\newblock


\bibitem[Michalkova et~al\mbox{.}(2022)]%
        {Dominika2022information}
\bibfield{author}{\bibinfo{person}{Dominika Michalkova}, \bibinfo{person}{Mario Parra-Rodriguez}, {and} \bibinfo{person}{Yashar Moshfeghi}.} \bibinfo{year}{2022}\natexlab{}.
\newblock \showarticletitle{Information Need Awareness: An EEG Study}. In \bibinfo{booktitle}{\emph{Proceedings of the 45th International ACM SIGIR Conference on Research and Development in Information Retrieval}} (Madrid, Spain) \emph{(\bibinfo{series}{SIGIR '22})}. \bibinfo{publisher}{Association for Computing Machinery}, \bibinfo{address}{New York, NY, USA}, \bibinfo{pages}{610–621}.
\newblock
\showISBNx{9781450387323}
\urldef\tempurl%
\url{https://doi.org/10.1145/3477495.3531999}
\showDOI{\tempurl}


\bibitem[Minas et~al\mbox{.}(2014)]%
        {randall2014putting}
\bibfield{author}{\bibinfo{person}{Randall~K. Minas}, \bibinfo{person}{Robert~F. Potter}, \bibinfo{person}{Alan~R. Dennis}, \bibinfo{person}{Valerie Bartelt}, {and} \bibinfo{person}{Soyoung Bae}.} \bibinfo{year}{2014}\natexlab{}.
\newblock \showarticletitle{Putting on the Thinking Cap: Using NeuroIS to Understand Information Processing Biases in Virtual Teams}.
\newblock \bibinfo{journal}{\emph{Journal of Management Information Systems}} \bibinfo{volume}{30}, \bibinfo{number}{4} (\bibinfo{year}{2014}), \bibinfo{pages}{49--82}.
\newblock
\urldef\tempurl%
\url{https://doi.org/10.2753/MIS0742-1222300403}
\showDOI{\tempurl}
\showeprint{https://doi.org/10.2753/MIS0742-1222300403}


\bibitem[Mohamed et~al\mbox{.}(2018)]%
        {mohamed2018characterizing}
\bibfield{author}{\bibinfo{person}{Zainab Mohamed}, \bibinfo{person}{Mohamed El~Halaby}, \bibinfo{person}{Tamer Said}, \bibinfo{person}{Doaa Shawky}, {and} \bibinfo{person}{Ashraf Badawi}.} \bibinfo{year}{2018}\natexlab{}.
\newblock \showarticletitle{Characterizing focused attention and working memory using EEG}.
\newblock \bibinfo{journal}{\emph{Sensors}} \bibinfo{volume}{18}, \bibinfo{number}{11} (\bibinfo{year}{2018}), \bibinfo{pages}{3743}.
\newblock


\bibitem[Morales and Bowers(2022)]%
        {MORALES2022101067}
\bibfield{author}{\bibinfo{person}{Santiago Morales} {and} \bibinfo{person}{Maureen~E. Bowers}.} \bibinfo{year}{2022}\natexlab{}.
\newblock \showarticletitle{Time-frequency analysis methods and their application in developmental EEG data}.
\newblock \bibinfo{journal}{\emph{Developmental Cognitive Neuroscience}}  \bibinfo{volume}{54} (\bibinfo{year}{2022}), \bibinfo{pages}{101067}.
\newblock
\showISSN{1878-9293}
\urldef\tempurl%
\url{https://doi.org/10.1016/j.dcn.2022.101067}
\showDOI{\tempurl}


\bibitem[Moravec et~al\mbox{.}(2018)]%
        {moravec2018fake}
\bibfield{author}{\bibinfo{person}{Patricia Moravec}, \bibinfo{person}{Randall Minas}, {and} \bibinfo{person}{Alan~R Dennis}.} \bibinfo{year}{2018}\natexlab{}.
\newblock \showarticletitle{Fake news on social media: People believe what they want to believe when it makes no sense at all}.
\newblock \bibinfo{journal}{\emph{Kelley School of Business research paper}} \bibinfo{number}{18-87} (\bibinfo{year}{2018}).
\newblock


\bibitem[Morgan(2009)]%
        {morgan2009picture}
\bibfield{author}{\bibinfo{person}{Hani Morgan}.} \bibinfo{year}{2009}\natexlab{}.
\newblock \showarticletitle{Picture book biographies for young children: A way to teach multiple perspectives}.
\newblock \bibinfo{journal}{\emph{Early Childhood Education Journal}}  \bibinfo{volume}{37} (\bibinfo{year}{2009}), \bibinfo{pages}{219--227}.
\newblock


\bibitem[Moshfeghi and Jose(2013)]%
        {moshfeghi2013effective}
\bibfield{author}{\bibinfo{person}{Yashar Moshfeghi} {and} \bibinfo{person}{Joemon~M Jose}.} \bibinfo{year}{2013}\natexlab{}.
\newblock \showarticletitle{An effective implicit relevance feedback technique using affective, physiological and behavioural features}. In \bibinfo{booktitle}{\emph{Proceedings of the 36th international ACM SIGIR conference on Research and development in information retrieval}}. \bibinfo{pages}{133--142}.
\newblock


\bibitem[Moshfeghi et~al\mbox{.}(2019)]%
        {moshfeghi2019towards}
\bibfield{author}{\bibinfo{person}{Yashar Moshfeghi}, \bibinfo{person}{Peter Triantafillou}, {and} \bibinfo{person}{Frank Pollick}.} \bibinfo{year}{2019}\natexlab{}.
\newblock \showarticletitle{Towards predicting a realisation of an information need based on brain signals}. In \bibinfo{booktitle}{\emph{The world wide web conference}}. \bibinfo{pages}{1300--1309}.
\newblock


\bibitem[Muse(2023)]%
        {Muse2023}
\bibfield{author}{\bibinfo{person}{Muse}.} \bibinfo{year}{2023}\natexlab{}.
\newblock \bibinfo{booktitle}{\emph{Muse EEG-Powered Meditation \& Sleep Headband}}.
\newblock
\urldef\tempurl%
\url{https://choosemuse.com/}
\showURL{%
\tempurl}
\newblock
\shownote{Accessed: October 21, 2023}.


\bibitem[Nakamura et~al\mbox{.}(2017)]%
        {nakamura2017ear}
\bibfield{author}{\bibinfo{person}{Takashi Nakamura}, \bibinfo{person}{Valentin Goverdovsky}, {and} \bibinfo{person}{Danilo~P Mandic}.} \bibinfo{year}{2017}\natexlab{}.
\newblock \showarticletitle{In-ear EEG biometrics for feasible and readily collectable real-world person authentication}.
\newblock \bibinfo{journal}{\emph{IEEE Transactions on Information Forensics and Security}} \bibinfo{volume}{13}, \bibinfo{number}{3} (\bibinfo{year}{2017}), \bibinfo{pages}{648--661}.
\newblock


\bibitem[Nakano and Ishii(2010)]%
        {gazehumanagent2010}
\bibfield{author}{\bibinfo{person}{Yukiko~I. Nakano} {and} \bibinfo{person}{Ryo Ishii}.} \bibinfo{year}{2010}\natexlab{}.
\newblock \showarticletitle{Estimating User's Engagement from Eye-Gaze Behaviors in Human-Agent Conversations}. In \bibinfo{booktitle}{\emph{Proceedings of the 15th International Conference on Intelligent User Interfaces}} (Hong Kong, China) \emph{(\bibinfo{series}{IUI '10})}. \bibinfo{publisher}{Association for Computing Machinery}, \bibinfo{address}{New York, NY, USA}, \bibinfo{pages}{139–148}.
\newblock
\showISBNx{9781605585154}
\urldef\tempurl%
\url{https://doi.org/10.1145/1719970.1719990}
\showDOI{\tempurl}


\bibitem[Novin and Meyers(2017)]%
        {novin2017making}
\bibfield{author}{\bibinfo{person}{Alamir Novin} {and} \bibinfo{person}{Eric Meyers}.} \bibinfo{year}{2017}\natexlab{}.
\newblock \showarticletitle{Making sense of conflicting science information: Exploring bias in the search engine result page}. In \bibinfo{booktitle}{\emph{Proceedings of the 2017 conference on conference human information interaction and retrieval}}. \bibinfo{pages}{175--184}.
\newblock


\bibitem[Odede and Frommholz(2024)]%
        {julius2024jaybot}
\bibfield{author}{\bibinfo{person}{Julius Odede} {and} \bibinfo{person}{Ingo Frommholz}.} \bibinfo{year}{2024}\natexlab{}.
\newblock \showarticletitle{JayBot -- Aiding University Students and Admission with an LLM-based Chatbot}. In \bibinfo{booktitle}{\emph{Proceedings of the 2024 Conference on Human Information Interaction and Retrieval}} (<conf-loc>, <city>Sheffield</city>, <country>United Kingdom</country>, </conf-loc>) \emph{(\bibinfo{series}{CHIIR '24})}. \bibinfo{publisher}{Association for Computing Machinery}, \bibinfo{address}{New York, NY, USA}, \bibinfo{pages}{391–395}.
\newblock
\showISBNx{9798400704345}
\urldef\tempurl%
\url{https://doi.org/10.1145/3627508.3638293}
\showDOI{\tempurl}


\bibitem[Ooko et~al\mbox{.}(2011)]%
        {ooko2011estimating}
\bibfield{author}{\bibinfo{person}{Ryota Ooko}, \bibinfo{person}{Ryo Ishii}, {and} \bibinfo{person}{Yukiko~I Nakano}.} \bibinfo{year}{2011}\natexlab{}.
\newblock \showarticletitle{Estimating a user’s conversational engagement based on head pose information}. In \bibinfo{booktitle}{\emph{Intelligent Virtual Agents: 10th International Conference, IVA 2011, Reykjavik, Iceland, September 15-17, 2011. Proceedings 11}}. Springer, \bibinfo{pages}{262--268}.
\newblock


\bibitem[Pathiyan~Cherumanal(2022)]%
        {pathiyan2022fairness}
\bibfield{author}{\bibinfo{person}{Sachin Pathiyan~Cherumanal}.} \bibinfo{year}{2022}\natexlab{}.
\newblock \showarticletitle{Fairness-Aware Question Answering for Intelligent Assistants} \emph{(\bibinfo{series}{SIGIR '22})}. \bibinfo{publisher}{Association for Computing Machinery}, \bibinfo{address}{New York, NY, USA}, \bibinfo{pages}{3492}.
\newblock
\showISBNx{9781450387323}
\urldef\tempurl%
\url{https://doi.org/10.1145/3477495.3531682}
\showDOI{\tempurl}


\bibitem[Pathiyan~Cherumanal et~al\mbox{.}(2021)]%
        {pathiyan2021argfairness}
\bibfield{author}{\bibinfo{person}{Sachin Pathiyan~Cherumanal}, \bibinfo{person}{Damiano Spina}, \bibinfo{person}{Falk Scholer}, {and} \bibinfo{person}{W.~Bruce Croft}.} \bibinfo{year}{2021}\natexlab{}.
\newblock \showarticletitle{Evaluating Fairness in Argument Retrieval}. In \bibinfo{booktitle}{\emph{Proceedings of the 30th ACM International Conference on Information \& Knowledge Management}} (Virtual Event, Queensland, Australia) \emph{(\bibinfo{series}{CIKM '21})}. \bibinfo{publisher}{Association for Computing Machinery}, \bibinfo{address}{New York, NY, USA}, \bibinfo{pages}{3363–3367}.
\newblock
\showISBNx{9781450384469}
\urldef\tempurl%
\url{https://doi.org/10.1145/3459637.3482099}
\showDOI{\tempurl}


\bibitem[Pathiyan~Cherumanal et~al\mbox{.}(2024)]%
        {pathiyan2024walert}
\bibfield{author}{\bibinfo{person}{Sachin Pathiyan~Cherumanal}, \bibinfo{person}{Lin Tian}, \bibinfo{person}{Futoon~M. Abushaqra}, \bibinfo{person}{Angel~Felipe Magnoss\~{a}o~de Paula}, \bibinfo{person}{Kaixin Ji}, \bibinfo{person}{Halil Ali}, \bibinfo{person}{Danula Hettiachchi}, \bibinfo{person}{Johanne~R. Trippas}, \bibinfo{person}{Falk Scholer}, {and} \bibinfo{person}{Damiano Spina}.} \bibinfo{year}{2024}\natexlab{}.
\newblock \showarticletitle{Walert: Putting Conversational Information Seeking Knowledge into Action by Building and Evaluating a Large Language Model-Powered Chatbot}. In \bibinfo{booktitle}{\emph{Proceedings of the 2024 Conference on Human Information Interaction and Retrieval}} (<conf-loc>, <city>Sheffield</city>, <country>United Kingdom</country>, </conf-loc>) \emph{(\bibinfo{series}{CHIIR '24})}. \bibinfo{publisher}{Association for Computing Machinery}, \bibinfo{address}{New York, NY, USA}, \bibinfo{pages}{401–405}.
\newblock
\showISBNx{9798400704345}
\urldef\tempurl%
\url{https://doi.org/10.1145/3627508.3638309}
\showDOI{\tempurl}


\bibitem[Pham et~al\mbox{.}(2021)]%
        {pham2021heart}
\bibfield{author}{\bibinfo{person}{Tam Pham}, \bibinfo{person}{Zen~Juen Lau}, \bibinfo{person}{SH~Annabel Chen}, {and} \bibinfo{person}{Dominique Makowski}.} \bibinfo{year}{2021}\natexlab{}.
\newblock \showarticletitle{Heart rate variability in psychology: A review of HRV indices and an analysis tutorial}.
\newblock \bibinfo{journal}{\emph{Sensors}} \bibinfo{volume}{21}, \bibinfo{number}{12} (\bibinfo{year}{2021}), \bibinfo{pages}{3998}.
\newblock


\bibitem[Picou et~al\mbox{.}(2011)]%
        {picou2011visual}
\bibfield{author}{\bibinfo{person}{Erin~M. Picou}, \bibinfo{person}{Todd~A. Ricketts}, {and} \bibinfo{person}{Benjamin~WY. Hornsby}.} \bibinfo{year}{2011}\natexlab{}.
\newblock \showarticletitle{Visual cues and listening effort: Individual variability}.
\newblock \bibinfo{journal}{\emph{Journal of Speech, Language, and Hearing Research}} \bibinfo{volume}{54}, \bibinfo{number}{5} (\bibinfo{year}{2011}).
\newblock


\bibitem[Ploger et~al\mbox{.}(2021)]%
        {ploger2021psychophysiological}
\bibfield{author}{\bibinfo{person}{Gavin~W Ploger}, \bibinfo{person}{Johnanna Dunaway}, \bibinfo{person}{Patrick Fournier}, {and} \bibinfo{person}{Stuart Soroka}.} \bibinfo{year}{2021}\natexlab{}.
\newblock \showarticletitle{The psychophysiological correlates of cognitive dissonance}.
\newblock \bibinfo{journal}{\emph{Politics and the Life Sciences}} \bibinfo{volume}{40}, \bibinfo{number}{2} (\bibinfo{year}{2021}), \bibinfo{pages}{202--212}.
\newblock


\bibitem[Radlinski and Craswell(2017)]%
        {radlinski2017theoretical}
\bibfield{author}{\bibinfo{person}{Filip Radlinski} {and} \bibinfo{person}{Nick Craswell}.} \bibinfo{year}{2017}\natexlab{}.
\newblock \showarticletitle{A theoretical framework for conversational search}. In \bibinfo{booktitle}{\emph{Proceedings of the 2017 conference on conference human information interaction and retrieval}}. \bibinfo{pages}{117--126}.
\newblock


\bibitem[Rainey et~al\mbox{.}(2020)]%
        {rainey2020brain}
\bibfield{author}{\bibinfo{person}{Stephen Rainey}, \bibinfo{person}{St{\'e}phanie Martin}, \bibinfo{person}{Andy Christen}, \bibinfo{person}{Pierre M{\'e}gevand}, {and} \bibinfo{person}{Eric Fourneret}.} \bibinfo{year}{2020}\natexlab{}.
\newblock \showarticletitle{Brain recording, mind-reading, and neurotechnology: ethical issues from consumer devices to brain-based speech decoding}.
\newblock \bibinfo{journal}{\emph{Science and engineering ethics}}  \bibinfo{volume}{26} (\bibinfo{year}{2020}), \bibinfo{pages}{2295--2311}.
\newblock


\bibitem[Regev et~al\mbox{.}(2019)]%
        {regev2019propagation}
\bibfield{author}{\bibinfo{person}{Mor Regev}, \bibinfo{person}{Erez Simony}, \bibinfo{person}{Katherine Lee}, \bibinfo{person}{Kean~Ming Tan}, \bibinfo{person}{Janice Chen}, {and} \bibinfo{person}{Uri Hasson}.} \bibinfo{year}{2019}\natexlab{}.
\newblock \showarticletitle{Propagation of information along the cortical hierarchy as a function of attention while reading and listening to stories}.
\newblock \bibinfo{journal}{\emph{Cerebral Cortex}} \bibinfo{volume}{29}, \bibinfo{number}{10} (\bibinfo{year}{2019}), \bibinfo{pages}{4017--4034}.
\newblock


\bibitem[Richter and Maier(2017)]%
        {richter2017comprehension}
\bibfield{author}{\bibinfo{person}{Tobias Richter} {and} \bibinfo{person}{Johanna Maier}.} \bibinfo{year}{2017}\natexlab{}.
\newblock \showarticletitle{Comprehension of multiple documents with conflicting information: A two-step model of validation}.
\newblock \bibinfo{journal}{\emph{Educational Psychologist}} \bibinfo{volume}{52}, \bibinfo{number}{3} (\bibinfo{year}{2017}), \bibinfo{pages}{148--166}.
\newblock


\bibitem[Riedl et~al\mbox{.}(2014)]%
        {riedl2014towards}
\bibfield{author}{\bibinfo{person}{Ren{\'e} Riedl}, \bibinfo{person}{Fred~D Davis}, {and} \bibinfo{person}{Alan~R Hevner}.} \bibinfo{year}{2014}\natexlab{}.
\newblock \showarticletitle{Towards a NeuroIS research methodology: intensifying the discussion on methods, tools, and measurement}.
\newblock \bibinfo{journal}{\emph{Journal of the Association for Information Systems}} \bibinfo{volume}{15}, \bibinfo{number}{10} (\bibinfo{year}{2014}), \bibinfo{pages}{4}.
\newblock


\bibitem[Sa and Yuan(2020)]%
        {sa2020examining}
\bibfield{author}{\bibinfo{person}{Ning Sa} {and} \bibinfo{person}{Xiaojun Yuan}.} \bibinfo{year}{2020}\natexlab{}.
\newblock \showarticletitle{Examining users' partial query modification patterns in voice search}.
\newblock \bibinfo{journal}{\emph{Journal of the Association for Information Science and Technology}} \bibinfo{volume}{71}, \bibinfo{number}{3} (\bibinfo{year}{2020}), \bibinfo{pages}{251--263}.
\newblock


\bibitem[Schmidt et~al\mbox{.}(2019)]%
        {schmidt2019wearable}
\bibfield{author}{\bibinfo{person}{Philip Schmidt}, \bibinfo{person}{Attila Reiss}, \bibinfo{person}{Robert D{\"u}richen}, {and} \bibinfo{person}{Kristof Van~Laerhoven}.} \bibinfo{year}{2019}\natexlab{}.
\newblock \showarticletitle{Wearable-based affect recognition—A review}.
\newblock \bibinfo{journal}{\emph{Sensors}} \bibinfo{volume}{19}, \bibinfo{number}{19} (\bibinfo{year}{2019}), \bibinfo{pages}{4079}.
\newblock


\bibitem[Schneegass et~al\mbox{.}(2023)]%
        {schneegass2023future}
\bibfield{author}{\bibinfo{person}{Christina Schneegass}, \bibinfo{person}{Max~L Wilson}, \bibinfo{person}{Horia~A. Maior}, \bibinfo{person}{Francesco Chiossi}, \bibinfo{person}{Anna~L Cox}, {and} \bibinfo{person}{Jason Wiese}.} \bibinfo{year}{2023}\natexlab{}.
\newblock \showarticletitle{The Future of Cognitive Personal Informatics}. In \bibinfo{booktitle}{\emph{Proceedings of the 25th International Conference on Mobile Human-Computer Interaction}} (Athens, Greece) \emph{(\bibinfo{series}{MobileHCI '23 Companion})}. \bibinfo{publisher}{Association for Computing Machinery}, \bibinfo{address}{New York, NY, USA}, Article \bibinfo{articleno}{35}, \bibinfo{numpages}{5}~pages.
\newblock
\showISBNx{9781450399241}
\urldef\tempurl%
\url{https://doi.org/10.1145/3565066.3609790}
\showDOI{\tempurl}


\bibitem[Schneiders(2020)]%
        {schneiders2020remains}
\bibfield{author}{\bibinfo{person}{Pascal Schneiders}.} \bibinfo{year}{2020}\natexlab{}.
\newblock \showarticletitle{What remains in mind? Effectiveness and efficiency of explainers at conveying information}.
\newblock \bibinfo{journal}{\emph{Media and Communication}} \bibinfo{volume}{8}, \bibinfo{number}{1} (\bibinfo{year}{2020}), \bibinfo{pages}{218--231}.
\newblock


\bibitem[Setlur and Tory(2022)]%
        {Setlur2022analyticalchatbot}
\bibfield{author}{\bibinfo{person}{Vidya Setlur} {and} \bibinfo{person}{Melanie Tory}.} \bibinfo{year}{2022}\natexlab{}.
\newblock \showarticletitle{How do you Converse with an Analytical Chatbot? Revisiting Gricean Maxims for Designing Analytical Conversational Behavior}. In \bibinfo{booktitle}{\emph{Proceedings of the 2022 CHI Conference on Human Factors in Computing Systems}} (New Orleans, LA, USA) \emph{(\bibinfo{series}{CHI '22})}. \bibinfo{publisher}{Association for Computing Machinery}, \bibinfo{address}{New York, NY, USA}, Article \bibinfo{articleno}{29}, \bibinfo{numpages}{17}~pages.
\newblock
\showISBNx{9781450391573}
\urldef\tempurl%
\url{https://doi.org/10.1145/3491102.3501972}
\showDOI{\tempurl}


\bibitem[Shah and Bender(2022)]%
        {shah2022situating}
\bibfield{author}{\bibinfo{person}{Chirag Shah} {and} \bibinfo{person}{Emily~M. Bender}.} \bibinfo{year}{2022}\natexlab{}.
\newblock \showarticletitle{Situating Search}. In \bibinfo{booktitle}{\emph{Proceedings of the 2022 Conference on Human Information Interaction and Retrieval}} (Regensburg, Germany) \emph{(\bibinfo{series}{CHIIR '22})}. \bibinfo{publisher}{Association for Computing Machinery}, \bibinfo{address}{New York, NY, USA}, \bibinfo{pages}{221–232}.
\newblock
\showISBNx{9781450391863}
\urldef\tempurl%
\url{https://doi.org/10.1145/3498366.3505816}
\showDOI{\tempurl}


\bibitem[Sharma et~al\mbox{.}(2024)]%
        {sharma2024echo}
\bibfield{author}{\bibinfo{person}{Nikhil Sharma}, \bibinfo{person}{Q.~Vera Liao}, {and} \bibinfo{person}{Ziang Xiao}.} \bibinfo{year}{2024}\natexlab{}.
\newblock \showarticletitle{Generative Echo Chamber? Effects of LLM-Powered Search Systems on Diverse Information Seeking}. In \bibinfo{booktitle}{\emph{Proceedings of the 2024 CHI Conference on Human Factors in Computing Systems}} (Honolulu, HI, USA) \emph{(\bibinfo{series}{CHI ’24})}. \bibinfo{publisher}{Association for Computing Machinery}, \bibinfo{address}{New York, NY, USA}, \bibinfo{numpages}{17}~pages.
\newblock
\urldef\tempurl%
\url{https://doi.org/10.1145/3613904.3642459}
\showDOI{\tempurl}


\bibitem[Sitbon et~al\mbox{.}(2023)]%
        {sitbon2023neurodiverseperspective}
\bibfield{author}{\bibinfo{person}{Laurianne Sitbon}, \bibinfo{person}{Gerd Berget}, {and} \bibinfo{person}{Margot Brereton}.} \bibinfo{year}{2023}\natexlab{}.
\newblock \showarticletitle{Perspectives of Neurodiverse Participants in Interactive Information Retrieval}.
\newblock \bibinfo{journal}{\emph{Found. Trends Inf. Retr.}} \bibinfo{volume}{17}, \bibinfo{number}{2} (\bibinfo{date}{jul} \bibinfo{year}{2023}), \bibinfo{pages}{124–243}.
\newblock
\showISSN{1554-0669}
\urldef\tempurl%
\url{https://doi.org/10.1561/1500000086}
\showDOI{\tempurl}


\bibitem[Smirnova(2020)]%
        {smirova2020word}
\bibfield{author}{\bibinfo{person}{Anastasia Smirnova}.} \bibinfo{year}{2020}\natexlab{}.
\newblock \showarticletitle{Word Order Communicates User Intent in Search Queries}. In \bibinfo{booktitle}{\emph{Extended Abstracts of the 2020 CHI Conference on Human Factors in Computing Systems}} (Honolulu, HI, USA) \emph{(\bibinfo{series}{CHI EA '20})}. \bibinfo{publisher}{Association for Computing Machinery}, \bibinfo{address}{New York, NY, USA}, \bibinfo{pages}{1–8}.
\newblock
\showISBNx{9781450368193}
\urldef\tempurl%
\url{https://doi.org/10.1145/3334480.3375207}
\showDOI{\tempurl}


\bibitem[Sokolov et~al\mbox{.}(2002)]%
        {sokolov2002orienting}
\bibfield{author}{\bibinfo{person}{Evgeny~N Sokolov}, \bibinfo{person}{John~A Spinks}, \bibinfo{person}{Risto N{\"a}{\"a}t{\"a}nen}, {and} \bibinfo{person}{Heikki Lyytinen}.} \bibinfo{year}{2002}\natexlab{}.
\newblock \bibinfo{booktitle}{\emph{The orienting response in information processing.}}
\newblock \bibinfo{publisher}{Lawrence Erlbaum Associates Publishers}.
\newblock


\bibitem[Soprano et~al\mbox{.}(2024)]%
        {soprano2024cognitive}
\bibfield{author}{\bibinfo{person}{Michael Soprano}, \bibinfo{person}{Kevin Roitero}, \bibinfo{person}{David~La Barbera}, \bibinfo{person}{Davide Ceolin}, \bibinfo{person}{Damiano Spina}, \bibinfo{person}{Gianluca Demartini}, {and} \bibinfo{person}{Stefano Mizzaro}.} \bibinfo{year}{2024}\natexlab{}.
\newblock \showarticletitle{Cognitive Biases in Fact-Checking and Their Countermeasures: A Review}.
\newblock \bibinfo{journal}{\emph{Information Processing \& Management}} \bibinfo{volume}{61}, \bibinfo{number}{3} (\bibinfo{year}{2024}), \bibinfo{pages}{103672}.
\newblock
\showISSN{0306-4573}
\urldef\tempurl%
\url{https://doi.org/10.1016/j.ipm.2024.103672}
\showDOI{\tempurl}


\bibitem[Spina et~al\mbox{.}(2023)]%
        {spina2023human}
\bibfield{author}{\bibinfo{person}{Damiano Spina}, \bibinfo{person}{Mark Sanderson}, \bibinfo{person}{Daniel Angus}, \bibinfo{person}{Gianluca Demartini}, \bibinfo{person}{Dana Mckay}, \bibinfo{person}{Lauren~L. Saling}, {and} \bibinfo{person}{Ryen~W. White}.} \bibinfo{year}{2023}\natexlab{}.
\newblock \showarticletitle{Human-AI Cooperation to Tackle Misinformation and Polarization}.
\newblock \bibinfo{journal}{\emph{Commun. ACM}} \bibinfo{volume}{66}, \bibinfo{number}{7} (\bibinfo{date}{jun} \bibinfo{year}{2023}), \bibinfo{pages}{40–45}.
\newblock
\showISSN{0001-0782}
\urldef\tempurl%
\url{https://doi.org/10.1145/3588431}
\showDOI{\tempurl}


\bibitem[Sriram and Greenwald(2009)]%
        {sriram2009brief}
\bibfield{author}{\bibinfo{person}{Natarajan Sriram} {and} \bibinfo{person}{Anthony~G Greenwald}.} \bibinfo{year}{2009}\natexlab{}.
\newblock \showarticletitle{The Brief Implicit Association Test}.
\newblock \bibinfo{journal}{\emph{Experimental psychology}} \bibinfo{volume}{56}, \bibinfo{number}{4} (\bibinfo{year}{2009}), \bibinfo{pages}{283--294}.
\newblock


\bibitem[Suzuki and Yamamoto(2021)]%
        {suzuki2021characterizing}
\bibfield{author}{\bibinfo{person}{Masaki Suzuki} {and} \bibinfo{person}{Yusuke Yamamoto}.} \bibinfo{year}{2021}\natexlab{}.
\newblock \showarticletitle{Characterizing the Influence of Confirmation Bias on Web Search Behavior}.
\newblock \bibinfo{journal}{\emph{Frontiers in Psychology}}  \bibinfo{volume}{12} (\bibinfo{year}{2021}), \bibinfo{pages}{771948}.
\newblock


\bibitem[Sweller(2011)]%
        {SWELLER201137}
\bibfield{author}{\bibinfo{person}{John Sweller}.} \bibinfo{year}{2011}\natexlab{}.
\newblock \showarticletitle{CHAPTER TWO - Cognitive Load Theory}.
\newblock \bibinfo{series}{Psychology of Learning and Motivation}, Vol.~\bibinfo{volume}{55}. \bibinfo{publisher}{Academic Press}, \bibinfo{pages}{37--76}.
\newblock
\showISSN{0079-7421}
\urldef\tempurl%
\url{https://doi.org/10.1016/B978-0-12-387691-1.00002-8}
\showDOI{\tempurl}


\bibitem[Tabassum et~al\mbox{.}(2020)]%
        {Madiha2020alwayslisteningva}
\bibfield{author}{\bibinfo{person}{Madiha Tabassum}, \bibinfo{person}{Tomasz Kosi\'{n}ski}, \bibinfo{person}{Alisa Frik}, \bibinfo{person}{Nathan Malkin}, \bibinfo{person}{Primal Wijesekera}, \bibinfo{person}{Serge Egelman}, {and} \bibinfo{person}{Heather~Richter Lipford}.} \bibinfo{year}{2020}\natexlab{}.
\newblock \showarticletitle{Investigating Users' Preferences and Expectations for Always-Listening Voice Assistants}.
\newblock \bibinfo{journal}{\emph{Proc. ACM Interact. Mob. Wearable Ubiquitous Technol.}} \bibinfo{volume}{3}, \bibinfo{number}{4}, Article \bibinfo{articleno}{153} (\bibinfo{date}{sep} \bibinfo{year}{2020}), \bibinfo{numpages}{23}~pages.
\newblock
\urldef\tempurl%
\url{https://doi.org/10.1145/3369807}
\showDOI{\tempurl}


\bibitem[Trippas et~al\mbox{.}(2018)]%
        {trippas2018informing}
\bibfield{author}{\bibinfo{person}{Johanne~R Trippas}, \bibinfo{person}{Damiano Spina}, \bibinfo{person}{Lawrence Cavedon}, \bibinfo{person}{Hideo Joho}, {and} \bibinfo{person}{Mark Sanderson}.} \bibinfo{year}{2018}\natexlab{}.
\newblock \showarticletitle{Informing the design of spoken conversational search: Perspective paper}. In \bibinfo{booktitle}{\emph{Proceedings of the 2018 conference on human information interaction \& retrieval}}. \bibinfo{pages}{32--41}.
\newblock


\bibitem[Trippas et~al\mbox{.}(2017)]%
        {trippas2017people}
\bibfield{author}{\bibinfo{person}{Johanne~R Trippas}, \bibinfo{person}{Damiano Spina}, \bibinfo{person}{Lawrence Cavedon}, {and} \bibinfo{person}{Mark Sanderson}.} \bibinfo{year}{2017}\natexlab{}.
\newblock \showarticletitle{How do people interact in conversational speech-only search tasks: A preliminary analysis}. In \bibinfo{booktitle}{\emph{Proceedings of the 2017 conference on conference human information interaction and retrieval}}. \bibinfo{pages}{325--328}.
\newblock


\bibitem[Trippas et~al\mbox{.}(2015)]%
        {trippas2015resultspresentation}
\bibfield{author}{\bibinfo{person}{Johanne~R. Trippas}, \bibinfo{person}{Damiano Spina}, \bibinfo{person}{Mark Sanderson}, {and} \bibinfo{person}{Lawrence Cavedon}.} \bibinfo{year}{2015}\natexlab{}.
\newblock \showarticletitle{Results Presentation Methods for a Spoken Conversational Search System}. In \bibinfo{booktitle}{\emph{Proceedings of the First International Workshop on Novel Web Search Interfaces and Systems}} (Melbourne, Australia) \emph{(\bibinfo{series}{NWSearch '15})}. \bibinfo{publisher}{Association for Computing Machinery}, \bibinfo{address}{New York, NY, USA}, \bibinfo{pages}{13–15}.
\newblock
\showISBNx{9781450337892}
\urldef\tempurl%
\url{https://doi.org/10.1145/2810355.2810356}
\showDOI{\tempurl}


\bibitem[Trippas et~al\mbox{.}(2020)]%
        {trippas2020towards}
\bibfield{author}{\bibinfo{person}{Johanne~R. Trippas}, \bibinfo{person}{Damiano Spina}, \bibinfo{person}{Paul Thomas}, \bibinfo{person}{Mark Sanderson}, \bibinfo{person}{Hideo Joho}, {and} \bibinfo{person}{Lawrence Cavedon}.} \bibinfo{year}{2020}\natexlab{}.
\newblock \showarticletitle{Towards a Model for Spoken Conversational Search}.
\newblock \bibinfo{journal}{\emph{Information Processing \& Management}} \bibinfo{volume}{57}, \bibinfo{number}{2} (\bibinfo{year}{2020}), \bibinfo{pages}{102162}.
\newblock
\showISSN{0306-4573}
\urldef\tempurl%
\url{https://doi.org/10.1016/j.ipm.2019.102162}
\showDOI{\tempurl}


\bibitem[Vtyurina et~al\mbox{.}(2017)]%
        {vtyurina2017exploring}
\bibfield{author}{\bibinfo{person}{Alexandra Vtyurina}, \bibinfo{person}{Denis Savenkov}, \bibinfo{person}{Eugene Agichtein}, {and} \bibinfo{person}{Charles~LA Clarke}.} \bibinfo{year}{2017}\natexlab{}.
\newblock \showarticletitle{Exploring conversational search with humans, assistants, and wizards}. In \bibinfo{booktitle}{\emph{Proceedings of the 2017 chi conference extended abstracts on human factors in computing systems}}. \bibinfo{pages}{2187--2193}.
\newblock


\bibitem[Wall et~al\mbox{.}(2017)]%
        {wall2017warning}
\bibfield{author}{\bibinfo{person}{Emily Wall}, \bibinfo{person}{Leslie~M Blaha}, \bibinfo{person}{Lyndsey Franklin}, {and} \bibinfo{person}{Alex Endert}.} \bibinfo{year}{2017}\natexlab{}.
\newblock \showarticletitle{Warning, bias may occur: A proposed approach to detecting cognitive bias in interactive visual analytics}. In \bibinfo{booktitle}{\emph{2017 ieee conference on visual analytics science and technology (vast)}}. IEEE, \bibinfo{pages}{104--115}.
\newblock


\bibitem[Wei et~al\mbox{.}(2022)]%
        {wei2022esm}
\bibfield{author}{\bibinfo{person}{Jing Wei}, \bibinfo{person}{Benjamin Tag}, \bibinfo{person}{Johanne~R Trippas}, \bibinfo{person}{Tilman Dingler}, {and} \bibinfo{person}{Vassilis Kostakos}.} \bibinfo{year}{2022}\natexlab{}.
\newblock \showarticletitle{What Could Possibly Go Wrong When Interacting with Proactive Smart Speakers? A Case Study Using an ESM Application}. In \bibinfo{booktitle}{\emph{Proceedings of the 2022 CHI Conference on Human Factors in Computing Systems}} (New Orleans, LA, USA) \emph{(\bibinfo{series}{CHI '22})}. \bibinfo{publisher}{Association for Computing Machinery}, \bibinfo{address}{New York, NY, USA}, Article \bibinfo{articleno}{276}, \bibinfo{numpages}{15}~pages.
\newblock
\showISBNx{9781450391573}
\urldef\tempurl%
\url{https://doi.org/10.1145/3491102.3517432}
\showDOI{\tempurl}


\bibitem[White(2013)]%
        {belief2013white}
\bibfield{author}{\bibinfo{person}{Ryen White}.} \bibinfo{year}{2013}\natexlab{}.
\newblock \showarticletitle{Beliefs and Biases in Web Search}. In \bibinfo{booktitle}{\emph{Proceedings of the 36th International ACM SIGIR Conference on Research and Development in Information Retrieval}} (Dublin, Ireland) \emph{(\bibinfo{series}{SIGIR '13})}. \bibinfo{publisher}{Association for Computing Machinery}, \bibinfo{address}{New York, NY, USA}, \bibinfo{pages}{3–12}.
\newblock
\showISBNx{9781450320344}
\urldef\tempurl%
\url{https://doi.org/10.1145/2484028.2484053}
\showDOI{\tempurl}


\bibitem[Wilke and Mata(2012)]%
        {wilke2012cognitive}
\bibfield{author}{\bibinfo{person}{Andreas Wilke} {and} \bibinfo{person}{Rui Mata}.} \bibinfo{year}{2012}\natexlab{}.
\newblock \showarticletitle{Cognitive bias}.
\newblock In \bibinfo{booktitle}{\emph{Encyclopedia of human behavior}}. \bibinfo{publisher}{Academic Press}, \bibinfo{pages}{531--535}.
\newblock


\bibitem[Wilson et~al\mbox{.}(2021)]%
        {wilson2021objective}
\bibfield{author}{\bibinfo{person}{Justin~C. Wilson}, \bibinfo{person}{Suku Nair}, \bibinfo{person}{Sandro Scielzo}, {and} \bibinfo{person}{Eric~C. Larson}.} \bibinfo{year}{2021}\natexlab{}.
\newblock \showarticletitle{Objective Measures of Cognitive Load Using Deep Multi-Modal Learning: A Use-Case in Aviation}.
\newblock \bibinfo{journal}{\emph{Proc. ACM Interact. Mob. Wearable Ubiquitous Technol.}} \bibinfo{volume}{5}, \bibinfo{number}{1}, Article \bibinfo{articleno}{40} (\bibinfo{date}{mar} \bibinfo{year}{2021}), \bibinfo{numpages}{35}~pages.
\newblock
\urldef\tempurl%
\url{https://doi.org/10.1145/3448111}
\showDOI{\tempurl}


\bibitem[Yang et~al\mbox{.}(2023)]%
        {Kangning2023Multimodal}
\bibfield{author}{\bibinfo{person}{Kangning Yang}, \bibinfo{person}{Chaofan Wang}, \bibinfo{person}{Yue Gu}, \bibinfo{person}{Zhanna Sarsenbayeva}, \bibinfo{person}{Benjamin Tag}, \bibinfo{person}{Tilman Dingler}, \bibinfo{person}{Greg Wadley}, {and} \bibinfo{person}{Jorge Goncalves}.} \bibinfo{year}{2023}\natexlab{}.
\newblock \showarticletitle{Behavioral and Physiological Signals-Based Deep Multimodal Approach for Mobile Emotion Recognition}.
\newblock \bibinfo{journal}{\emph{IEEE Transactions on Affective Computing}} \bibinfo{volume}{14}, \bibinfo{number}{2} (\bibinfo{year}{2023}), \bibinfo{pages}{1082--1097}.
\newblock
\urldef\tempurl%
\url{https://doi.org/10.1109/TAFFC.2021.3100868}
\showDOI{\tempurl}


\bibitem[Ye et~al\mbox{.}(2022)]%
        {ye2022towards}
\bibfield{author}{\bibinfo{person}{Ziyi Ye}, \bibinfo{person}{Xiaohui Xie}, \bibinfo{person}{Yiqun Liu}, \bibinfo{person}{Zhihong Wang}, \bibinfo{person}{Xuesong Chen}, \bibinfo{person}{Min Zhang}, {and} \bibinfo{person}{Shaoping Ma}.} \bibinfo{year}{2022}\natexlab{}.
\newblock \showarticletitle{Towards a Better Understanding of Human Reading Comprehension with Brain Signals}. In \bibinfo{booktitle}{\emph{Proceedings of the ACM Web Conference 2022}} (Virtual Event, Lyon, France) \emph{(\bibinfo{series}{WWW '22})}. \bibinfo{publisher}{Association for Computing Machinery}, \bibinfo{address}{New York, NY, USA}, \bibinfo{pages}{380–391}.
\newblock
\showISBNx{9781450390965}
\urldef\tempurl%
\url{https://doi.org/10.1145/3485447.3511966}
\showDOI{\tempurl}


\bibitem[Yfantidou et~al\mbox{.}(2023)]%
        {Yfantidou2023bias}
\bibfield{author}{\bibinfo{person}{Sofia Yfantidou}, \bibinfo{person}{Pavlos Sermpezis}, \bibinfo{person}{Athena Vakali}, {and} \bibinfo{person}{Ricardo Baeza-Yates}.} \bibinfo{year}{2023}\natexlab{}.
\newblock \showarticletitle{Uncovering Bias in Personal Informatics}.
\newblock \bibinfo{journal}{\emph{Proc. ACM Interact. Mob. Wearable Ubiquitous Technol.}} \bibinfo{volume}{7}, \bibinfo{number}{3}, Article \bibinfo{articleno}{139} (\bibinfo{date}{sep} \bibinfo{year}{2023}), \bibinfo{numpages}{30}~pages.
\newblock
\urldef\tempurl%
\url{https://doi.org/10.1145/3610914}
\showDOI{\tempurl}


\bibitem[Yu and Han(2023)]%
        {yu2023people}
\bibfield{author}{\bibinfo{person}{Heeseung Yu} {and} \bibinfo{person}{Eunkyoung Han}.} \bibinfo{year}{2023}\natexlab{}.
\newblock \showarticletitle{People see what they want to see: an EEG study}.
\newblock \bibinfo{journal}{\emph{Cognitive Neurodynamics}} (\bibinfo{year}{2023}), \bibinfo{pages}{1--15}.
\newblock


\bibitem[Yunkaporta(2023)]%
        {yunkaporta2023right}
\bibfield{author}{\bibinfo{person}{Tyson Yunkaporta}.} \bibinfo{year}{2023}\natexlab{}.
\newblock \bibinfo{booktitle}{\emph{Right Story, Wrong Story: Adventures in Indigenous Thinking}}.
\newblock \bibinfo{publisher}{Text Publishing Company}.
\newblock
\showISBNx{9781922790439}


\bibitem[Yuste(2023)]%
        {yuste2023advocating}
\bibfield{author}{\bibinfo{person}{Rafael Yuste}.} \bibinfo{year}{2023}\natexlab{}.
\newblock \showarticletitle{Advocating for neurodata privacy and neurotechnology regulation}.
\newblock \bibinfo{journal}{\emph{Nature Protocols}} \bibinfo{volume}{18}, \bibinfo{number}{10} (\bibinfo{year}{2023}), \bibinfo{pages}{2869--2875}.
\newblock


\bibitem[Zamani et~al\mbox{.}(2023)]%
        {zamani2023conversational}
\bibfield{author}{\bibinfo{person}{Hamed Zamani}, \bibinfo{person}{Johanne~R Trippas}, \bibinfo{person}{Jeff Dalton}, \bibinfo{person}{Filip Radlinski}, {et~al\mbox{.}}} \bibinfo{year}{2023}\natexlab{}.
\newblock \showarticletitle{Conversational information seeking}.
\newblock \bibinfo{journal}{\emph{Foundations and Trends{\textregistered} in Information Retrieval}} \bibinfo{volume}{17}, \bibinfo{number}{3-4} (\bibinfo{year}{2023}), \bibinfo{pages}{244--456}.
\newblock


\bibitem[Zillich and Guenther(2021)]%
        {zillich2021selective}
\bibfield{author}{\bibinfo{person}{Arne~Freya Zillich} {and} \bibinfo{person}{Lars Guenther}.} \bibinfo{year}{2021}\natexlab{}.
\newblock \showarticletitle{Selective Exposure to Information on the Internet: Measuring Cognitive Dissonance and Selective Exposure with Eye-Tracking}.
\newblock \bibinfo{journal}{\emph{International Journal of Communication}}  \bibinfo{volume}{15} (\bibinfo{year}{2021}), \bibinfo{pages}{20}.
\newblock


\end{thebibliography}
\end{document}